\documentclass{article}

\usepackage{epsfig}
\usepackage{graphics}
\usepackage[left=2.5cm,top=3cm,right=2.5cm,bottom=3cm]{geometry}

\begin{document}


\title{PYRAMIR: Calibration and operation of a pyramid near-infrared wavefront
  sensor}
\date{}
\maketitle

{\bf D. Peter\footnotemark[1], M. Feldt\footnotemark[1], B. Dorner\footnotemark[1], T. Henning\footnotemark[1], S. Hippler\footnotemark[1], J.
  Aceituno\footnotemark[2]}\newline
\hspace*{0.6cm}{\footnotesize 1, Max-Planck-Institut f\"ur Astronomie, K\"onigstuhl 17, 69117
  Heidelberg, Germany } \newline
\hspace*{0.6cm}{\footnotesize 2, Centro Astron\'omico Hispanio Alem\'an, C/ Jes\'us Durb\'an
  Rem\'on, 2-2, 04004 Almeria, Spain}\vspace{1cm}\newline

\begin{abstract}
  The concept of pyramid wavefront sensors (PWFS) has been around about a
  decade by now. However there is still a great lack of characterizing
  measurements that allow the best operation of such a system under real life
  conditions at an astronomical telescope.\newline In this article we, therefore,
  investigate the behavior and robustness of the pyramid infrared wavefront
  sensor PYRAMIR mounted at the 3.5 m telescope at the Calar Alto Observatory
  under the influence of different error sources both intrinsic to the sensor,
  and arising in the preceding optical system. The intrinsic errors include
  diffraction effects on the pyramid edges and detector read out noise.
  \newline The external imperfections consist of a Gaussian profile in the
  intensity distribution in the pupil plane during calibration, the effect of
  an optically resolved reference source, and noncommon-path aberrations. We
  investigated the effect of three differently sized reference sources on the
  calibration of the PWFS.  For the noncommon-path aberrations the quality of
  the response of the system is quantified in terms of modal cross talk and
  aliasing.  We investigate the special behavior of the system regarding
  tip-tilt control. \newline
From our measurements we derive the method to optimize the
  calibration procedure and the setup of a PWFS adaptive optics (AO) system.
  We also calculate the total wavefront error arising from aliasing, modal
  cross talk, measurement error, and fitting error in order to optimize the
  number of calibrated modes for on-sky operations. These measurements result
  in a prediction of on-sky performance for various conditions.
\end{abstract}

\section{Introduction}\label{sec:Intr}
Within the context of AO systems pyramid wavefront sensors (PWFS) are a
relatively novel concept. The special interest in this concept arises from the
prediction of a gain in sensitivity -- and thus in limiting magnitude -- for a
nonmodulated PWFS over a Shack-Hartmann sensor (SHS) \cite{rag99} in closed-loop conditions for a
well-corrected point source. The definition of this gain is that in order to
achieve the same correction quality the PWFS needs less signal than the
SHS.  This gain in sensitivity results basically from
the fact that the accuracy of the measurement and the resulting reconstruction
error $\sigma^2_{SH}$ of the SHS depends on $\lambda\over{d_{sub}}$ with
$\lambda$ = sensing wavelength, $d_{sub}$ = subaperture size typically chosen
on the order of the Fried-parameter $r_0$ for typical site seeing at science
wave-length. In the case of the PWFS the measurement error $\sigma^2_{P}$
depends on $\lambda\over{D}$ with $D$ being the telescope diameter ($D >>
r_0$). From these formulas the difference in stellar magnitudes that are
needed to achieve the same Strehl ratio for the exemplary case of tip-tilt
only can be derived for both sensors as $\Delta m = -2.5
log\left({\sigma_P^2\over{\sigma_{SH}^2}}\right)\approx -2.5
log\left({r_0^2\over{D^2}}\right)$. Simulations show , for instance,
\cite{rag99} that the gain in sensitivity for a 4\,m class telescope and a
seeing of $0.5''$ ($r_0=20$\,cm) is predicted to be 2 magnitudes.
\subsection{The Pyramid Principle}\label{subsec:PP}
Wavefront sensing based on the pyramid principle has its origin in the
Foucault knife-edge test. The historical development of the PWFS is described
in \cite{Camp2006}.  
\begin{figure}[h!]
\centering
\includegraphics[width=12cm]{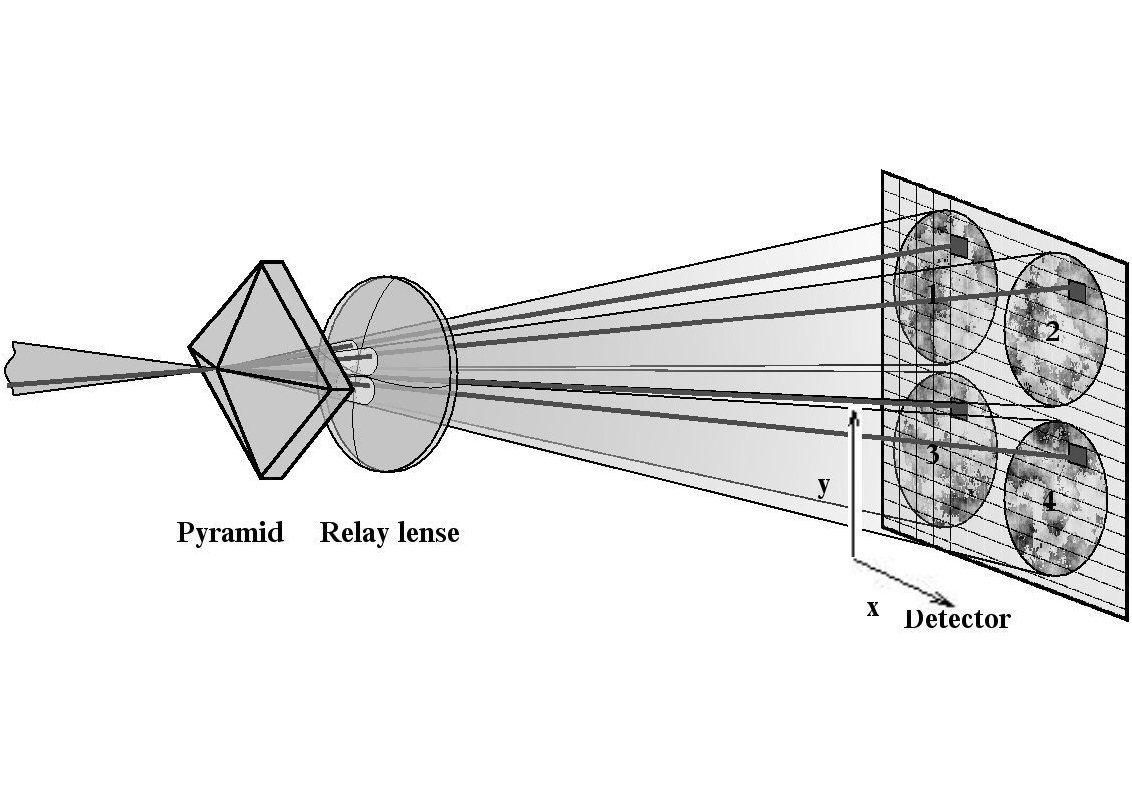}
  \caption{(Courtesy of S. Egner) The pyramid principle. An example of
    corresponding pixels is marked. The gradient in this position of the pupil
    is calculated from the intensity differences in between these pixels.}
  \label{im:f1}
\end{figure}
The optical setup of a pyramid sensor is shown in Figure
\ref{im:f1}.  The transmissive, four-sided pyramid prism is placed in the
focal plane.  The focus is placed on the tip of the
pyramid. After the
pyramid, a relay lens images the pupils onto the detector.\newline
The signal a four-sided AO PWFS system uses are the intensities inside four
pupil images. The illumination of these images depends on the aberrations of
the wavefront. The signal $S$ one extracts is the difference in intensities
$I_{1,2,3,4}$ between corresponding pixels in the four pupils ,i.e., the pixels
at the same optical position in the pupils as shown in Fig.\ref{im:f1}:

\begin{equation}S_x(x,y)={{I_1(x,y)+I_3(x,y)-[I_2(x,y)+I_4(x,y)]}\over{I_1(x,y)+I_2(x,y)+I_3(x,y)+I_4(x,y)}}\end{equation}
\begin{equation}S_y(x,y)={{I_1(x,y)+I_2(x,y)-[I_3(x,y)+I_4(x,y)]}\over{I_1(x,y)+I_2(x,y)+I_3(x,y)+I_4(x,y)}}\end{equation}
for the x- and y-direction, respectively. These signals $S$ are what we will refer
to as gradients, even if they might not exactly represent wavefront gradients.
In the limit of small perturbations and a telescope with infinite aperture the
frequency spectrum of the signal $S_x$ of our sensor is given by
\begin{equation}\widetilde{S_x}=i{\rm
    sgn}(u)\widetilde{\phi}(u,v).\end{equation}
Here $\widetilde{()}$ means Fourier transform, $\phi(u,v)$ the phase of the
electromagnetic wave with $u$ and $v$ the coordinates in Fourier space, and sgn the
sign-function (see \cite{Costa2003}). Thus the sensor is working as a
phase sensor in this regime.
\newline
The principle of the PWFS implies some limitations that either do not
occur in other sensor types, or have a much more "dramatic" impact on
PWFSs than on other sensor types.  In this class fall the structure of
the pyramid edges that cause both diffraction and
scattering, the read out noise of the system, the goodness of
centering the beam on the pyramid tip, noncommon-path aberrations,
and the homogeneity of the pupil illumination, the latter being
important especially during calibration.  It is these limitations,
that may ultimately influence the choice of this or another sensor
type, that will be examined in the course of this article.
\newline
The structure in the remaining part of the article is as follows: In the next section, the 
PYRAMIR system is described in detail. The optical path and the
possible detector read out modes are explained. Section 3 accounts for
the calibration procedure of the system. The peculiarities of tip-tilt
calibration and flattening the wavefront are presented.  Section 4
shows fundamental limitations to the performance of a pyramid system.
``Fundamental'' in this case is not meant to be based on natural laws
and constants, but on always-present aberrations and nonideal
conditions. Thus, the fundamental limitations discussed here can be
eased by careful alignment and set up of the system, but they can
never entirely be removed. In this context, we explore the effect of
static aberrations, different calibration light sources, and diffraction and
scattering on the pyramid edges to the response of the system.  
In the next section the implications of the modal cross talk of the system
aliasing and measurement error to the number of modes to be calibrated,
and the residual wavefront error are calculated. We end the section with a
prediction of the on-sky performance under different seeing conditions. 
The last section concludes the results of our
measurements and the implications for any pyramid wavefront sensor.\newline
On-sky results will be presented in a following paper \cite{Peter2008}.
\section{The PYRAMIR System}\label{sec:Overview}
PYRAMIR is a PWFS working in the effective wavelength regime between 1.26 and
2.4 $\mu$m (J,H,K-Bands). It is integrated into the ALFA-system at the 3.5m telescope of the Calar Alto observatory.\newline 
\begin{table}[h!]
   \caption{Properties of the PYRAMIR system}
   \label{hard}
   \centering

\begin{tabular}{c c}
  \hline\hline
  length & 1.1m\\ 
  diameter &0.206 m \\ 
  weight & 35 kg\\ 
  Detector temperature & 78 K \\ 
  Temp. stability & $\ge$ 24h \\ 
  mount & xyz-stage moved by stepper motors\\
  step size & 10$\mu$m  \\ 
  beam splitter & P90O10, P20O80, K/J \\ \hline
\end{tabular}
\end{table}
The properties of the
system are shown in Table \ref{hard}. The beam splitter can be exchanged
manually. Here, the choice is between 90\% light on PYRAMIR and 10\% on the
science camera (P90O10), 20\% on PYRAMIR and 80\% on the science camera
(P20O80), or K-band on PYRAMIR and J-band on the science camera.\newline 
\begin{figure}
\centering
  \includegraphics[width=12cm]{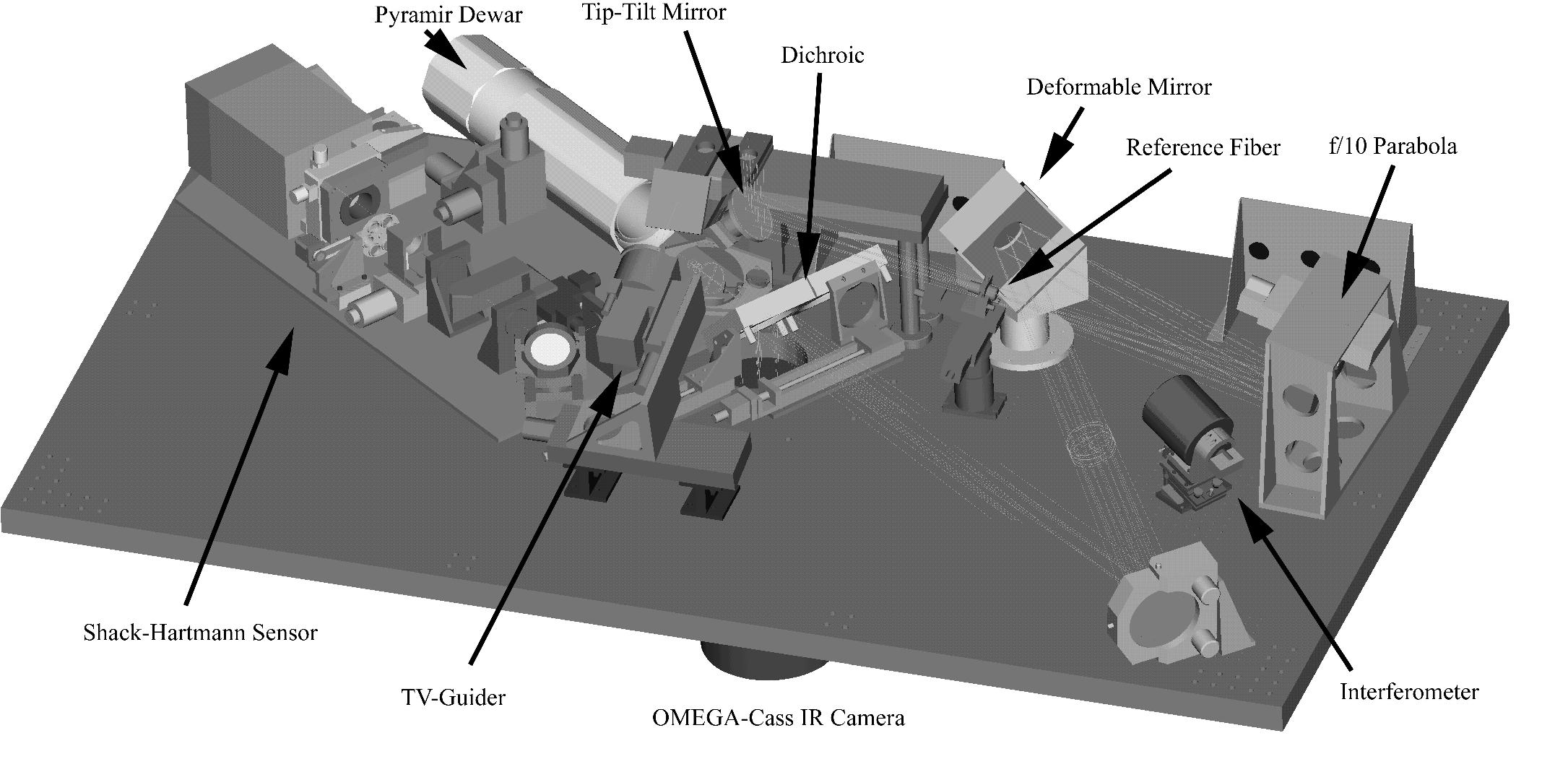}
  \caption{Setup of the adaptive optics bench ALFA.}
  \label{im:f2}
\end{figure}
The
optical path in the ALFA-system is shown in Fig.\ref{im:f2}. Coming from the
telescope the beam passes the mirror for tip-tilt (TT) correction. The
position of the telescope focus is downstream from the TT-mirror. At the
telescope's F/10 focal plane a calibration fiber can be introduced into the
beam. An off-axis parabola (F/10) is used to produce a collimated beam for the
illumination of the deformable mirror (DM, \begin{it}{Xinetics}\end{it} 97
actuator). The light is again focused by another off-axis parabola (F/25) and
then divided into the science and WFS-paths by the beam splitter of choice
(see table \ref{hard}).  
\begin{figure}
\centering
  \includegraphics[width=12cm]{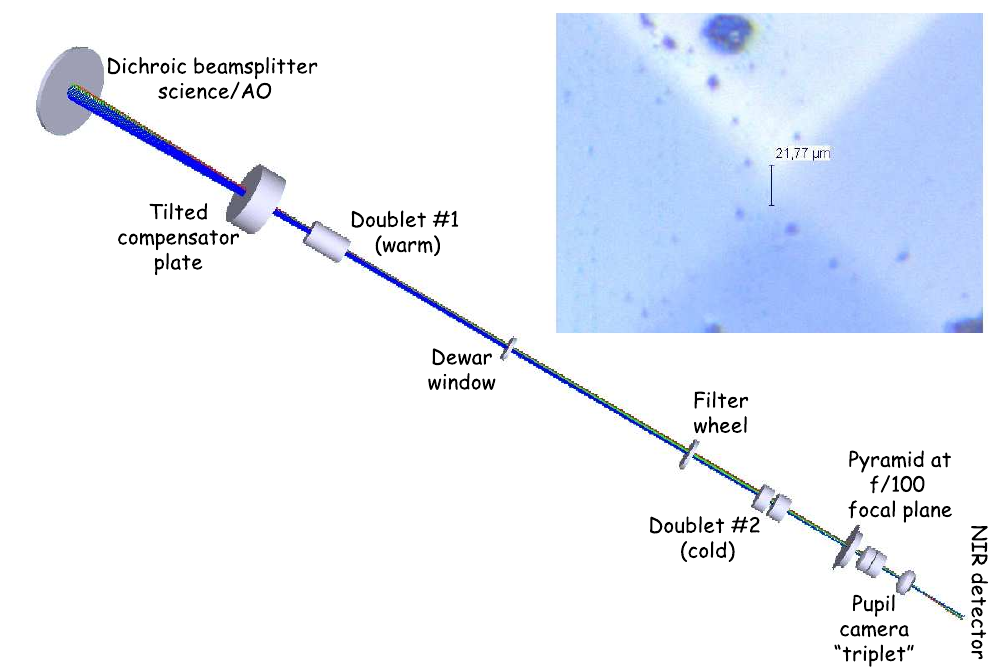}
  \caption{PYRAMIR arm of the system.}
  \label{im:f3}
\end{figure}
In the PYRAMIR arm (Fig.\ref{im:f3}) first a
compensator plate compensates the static and chromatic aberrations introduced
by the beam-splitter in the noncollimated beam. Then the light enters the
PYRAMIR system. A warm lens doublet together with a corresponding cold doublet
inside the dewar transforms the f/25 into a f/100 beam focused onto the tip of
the pyramid. The tip size of the pyramid is about 20 $\mu$m and the
diffraction-limited point-spread function (PSF) in the K-band has a full width at half-maximum (FWHM) of
$\approx$ 220 $\mu$m. The wavelength of correction can be chosen according to
Table \ref{detektor}. Additionally a spatial filter is introduced in front of
the pyramid to reduce aliasing effects, to minimize the sky background, and to
enable sensing on reasonably wide binaries. 
\begin{table}[h!]
  \caption{Properties of the PYRAMIR camera}
  \label{detektor}
  \centering
  \begin{tabular}{c c}\hline\hline
    Number of subapertures & 224 \\ 
    Detector & Hawaii I \\
    Pixel size & 18 $\mu$m \\
    maximum frame rate & $\approx$ 330 Hz \\
    frame size& 4 $\times$ 20x20 pixel \\
    field stops & 1'', 2'' \\
    Filters & J, H, K, H+K \\
    system gain & 3.8 $e^-$/ADU\\ 
    RON & 20 $e^-$ \\\hline
  \end{tabular}
\end{table}
The possible sizes of the field
stop are also given in Table \ref{detektor}. The final image is formed by a
lens-triplet onto the IR detector. The pupil diameter on the detector is about
320\,$\mu$m.  In order to place the detector in the pupil plane while keeping
the focus on the tip of the pyramid, one can move the detector in the
direction of the optical axis by about 2\,mm. The details of the PYRAMIR
detector are shown in table \ref{detektor}.

\subsection{Read out Modes}\label{subsec:Readout}
The detector, a Rockwell Hawaii I device, can be read out with three different
read out modes. Only two of these three modes are used during operation: the
line interlaced read (lir) and the multiple sampling read (msr). 
\begin{table}[h!]
  \caption{Read out modes of the PYRAMIR system}
  \label{read out}
  \centering
  \begin{tabular}{c c c c c c}\hline\hline
    name & Pattern & Integration Time & Maximum speed & Comments \\

    rrr &  reset-read-read & 50 \% & 165 Hz & lab use only \\ 
    lir & read-reset-read & 100 \% & 165 Hz & -- \\ 
    msr & reset-read-read-read-... & 100 \% & 330 Hz & loose frame \\ \hline
  \end{tabular}
\end{table}
For
specifications see Table \ref{read out}.\newline 
\begin{figure}[h!]
\centering
  \includegraphics[width=12cm]{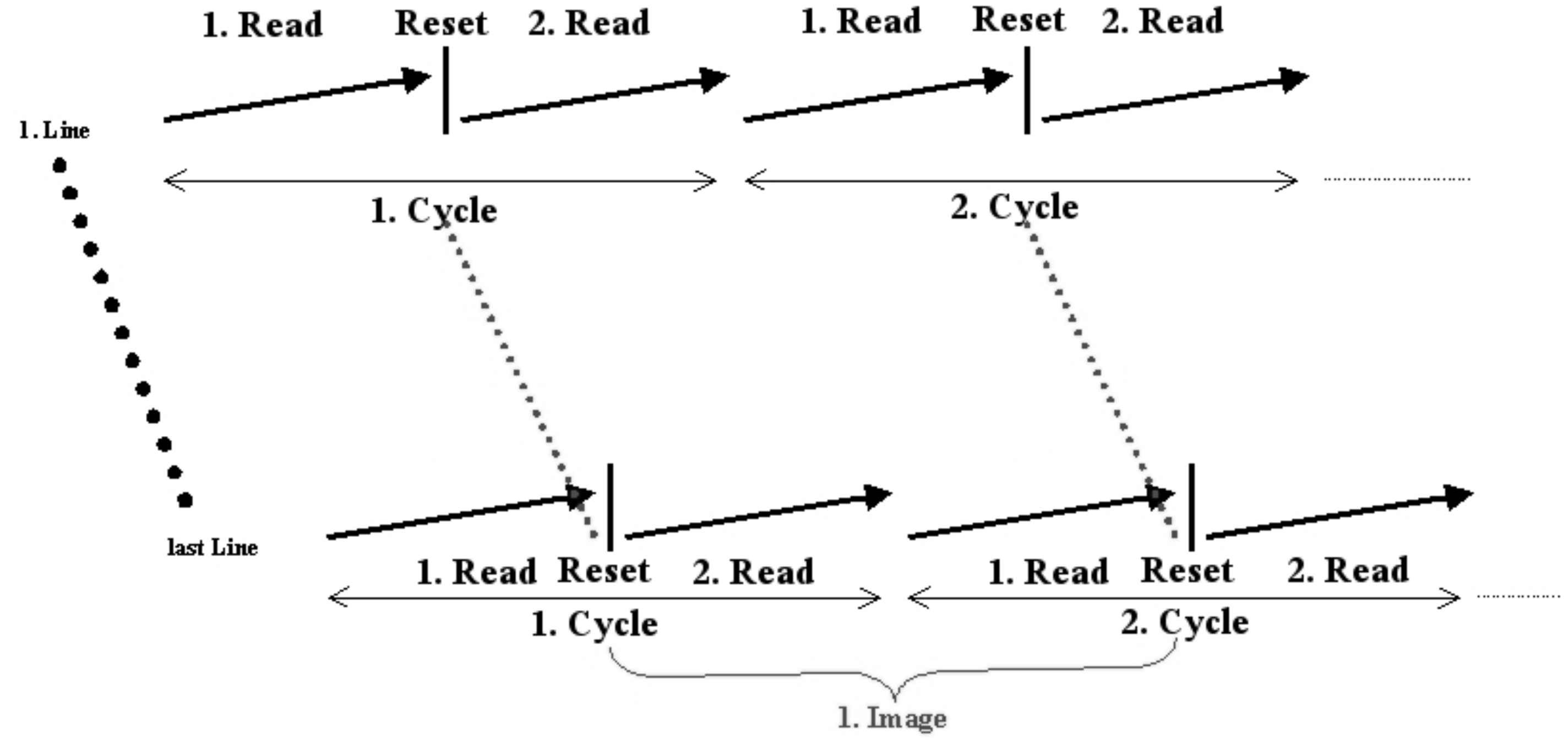}
  \caption{Read out pattern for the lir mode. Each line is read, reset, and
    read again before the next line is started. The image is built by the
    difference between the first frame of the second read and the second frame
    of the first read.}
  \label{im:f4}
\end{figure}
The lir mode is a
line-oriented mode with the read out pattern read-reset-read, that is
performed line by line. In this mode one line is read then reset and read
again. This procedure is then continued for the next line until the whole
array is read.  The integration time for the image is the time between the
second read of one pattern and the first read of the next (see
Fig.\ref{im:f4}). With this scheme, almost 100$\%$ of the cycle time can be
used as integration time (minus the reset time).\newline The msr mode is
frame-oriented, i.e., one resets the whole array, reads the whole array, reads
again the whole array etc. The pattern also differs from the lir mode. 
\begin{figure}
\centering
  \includegraphics[width=15cm]{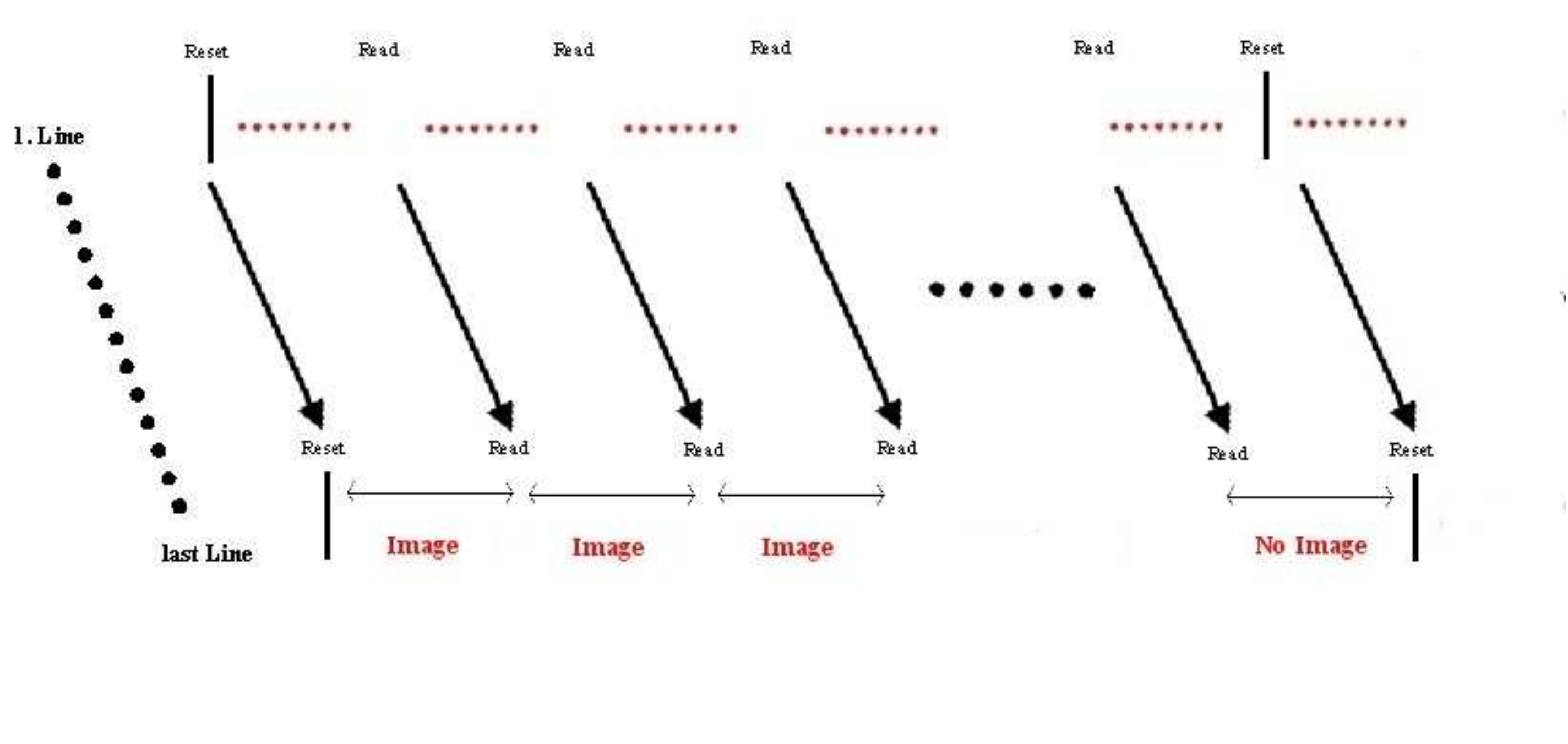}
  \caption{Read out pattern for the msr mode. The first image is
    constructed as the difference between the first read and a reset frame. The
    following images are built by the difference between two neighboring
    reads. The total signal on the detector is accumulated. The image between
    the second reset and the last read is lost.}
  \label{im:f5}
\end{figure}
First
the detector is reset and then one only reads out: reset-read-read-read-... as
shown in Fig.\ref{im:f5}. The signal is accumulated and the images used are
difference images between two neighboring read out frames.  Thus the images
$I_i$ are produced by resets $Res_i$ and reads $R_i$ as follows:
\begin{equation}I_1=R_1-Res_1,\,
  I_2=R_2-R_1,\,I_3=R_3-R_2,/,I_4=R_4-R_3...\end{equation}

In order not to saturate the detector while accumulating signal one has to
specify a maximum number of frames. For a star of fifth magnitude at 100 Hz
this maximum number is of the order of 100 frames ($\approx$ 1s integration).
The pattern runs up to the maximum frame and then starts again.  With this
pattern one will loose the frame at the end of every ramp : $Res_2-R_n$. Two
features are obvious here: first one can read out twice as fast as with the
lir mode because each image needs one read only, secondly already as mentioned
one looses the image between the last read and the reset of the following
pattern.  This means one performs one step with half the applied loop
frequency.

\section{Calibration} \label{sec:Calib}
\subsection{Tip-Tilt Calibration}\label{subsec:TTcalib}
Because the TT-mirror is located {\bf upstream} of the calibration light
source, the TT-modes and the high-order (HO) modes are calibrated differently.
For TT-calibration a 'star simulator', that mimics the light coming from the
telescope is used as light source.  This 'star simulator' is placed above the
entrance of the ALFA-system.  The 'star simulator' was originally introduced
for the visible wavefront sensor and is optimized for visible light.
Unfortunately, in the IR we face some strong static aberrations.  However,
these aberrations do not effect the TT-calibration. The basic effect is that
the pupils on the detector become smaller that introduces higher noise on the
measurement. The 'star simulator' cannot be used at the telescope. Therefore
our TT-calibration is done in the lab only.  It should be noted that care has
to be taken to achieve a good tip-tilt performance, because tip-tilt errors in
closed-loop (CL) will yield low light levels in at least one of the four
aperture images, thus affecting the limiting magnitude of the system (see
discussion in Sec.~\ref{subsec:Center}).  
\begin{figure}
\centering
  \includegraphics[width=12cm]{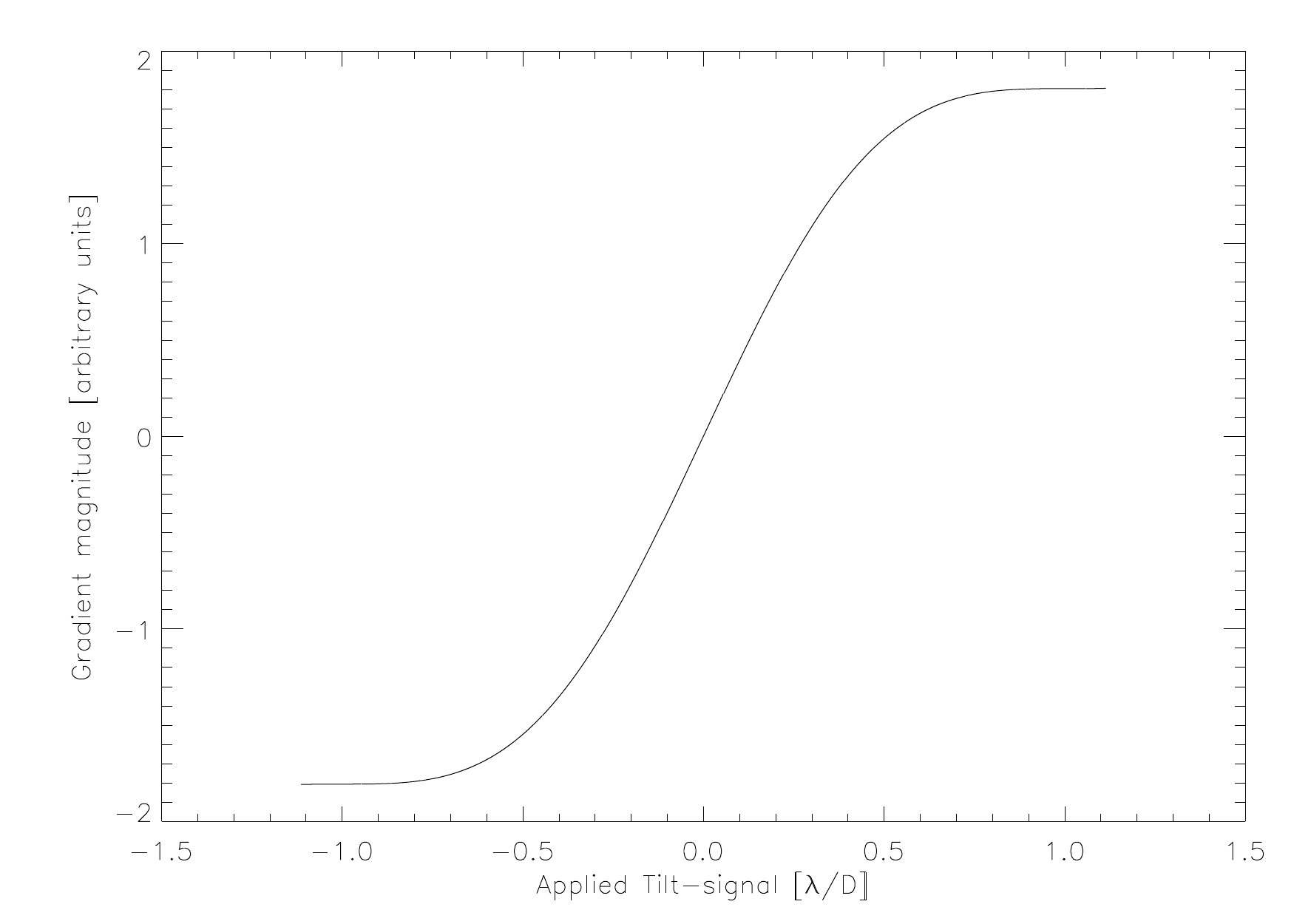}
  \caption{Theoretical Tilt signal. Shown is the length of the gradient
    vs. wavefront tilt in $\lambda/D$.  }
  \label{im:f6}
\end{figure}
Vice versa, a bad high-order
performance will clearly also affect tip-tilt correction, because the
effectively enlarged spot size flattens the curve shown in Fig.\ref{im:f6}.
Because the flattening occurs in only closed-loop when a
diffraction-limited spot was used during calibration, tip-tilt signals will always be measured
too small and correction will thus be slowed down, effectively lowering the
tip-tilt bandwidth.
\subsection{Phasing PYRAMIR}\label{subsec:phase}
The HO calibration is basically done with the usual procedure.  First we shape
the DM with the desired offset (dmBias) to guarantee a flat wavefront on
PYRAMIR in order to get the best possible calibration. To do so, the following
procedure is implemented. First the wavefront is flattened as much as possible
by applying aberrations to the DM and a check of the pupil images by ``eye.''
From this starting point a calibration of 20 modes is done. After the
calibration the mode offset is put to zero (that resembles not exactly a flat
wave but within an error of $\lambda/{200}$ root mean square (rms)). With
this offset we close the loop and flatten the wavefront on PYRAMIR.  This is
the starting point for the true calibration with the desired number of modes.
\section{Fundamental restrictions}\label{sec:restric}
There are fundamental properties of a PWFS system that can strongly influence
the performance of the system.  These properties are fundamental not so much
in the sense of natural laws and/or constants, but they can be mitigated by
appropriate design and alignment procedures.  However, they can never entirely
be removed! These are the diffraction and scattering effects on the pyramid
edges, a not-well centered beam on the pyramid tip, read out noise (RON) of the camera, a
nonhomogeneous illumination of the pupil during calibration, and
noncommon-path aberrations. The impact of each of these possible sources of
reduced performance is discussed below and measurements are presented that
show the effects of these error sources and the best way to manage them.
All these measurements are performed in open loop. However, they can be used
to quantify the performance of the system during CL operation. 
\subsection{Impact of the Pyramid Edges}\label{subsec:Diff}
There are two effects of the pyramid edges that yield a loss of light in the
pupil images. One is the diffraction at the edges even in the case of a
perfect pyramid. The second cause of light loss is the finite size of the
edges. Here, the light will be scattered anywhere, but it will not produce a
useful signal in any of the pupil images.\newline 
\begin{figure}[h!]
\centering
  \includegraphics[width=12cm]{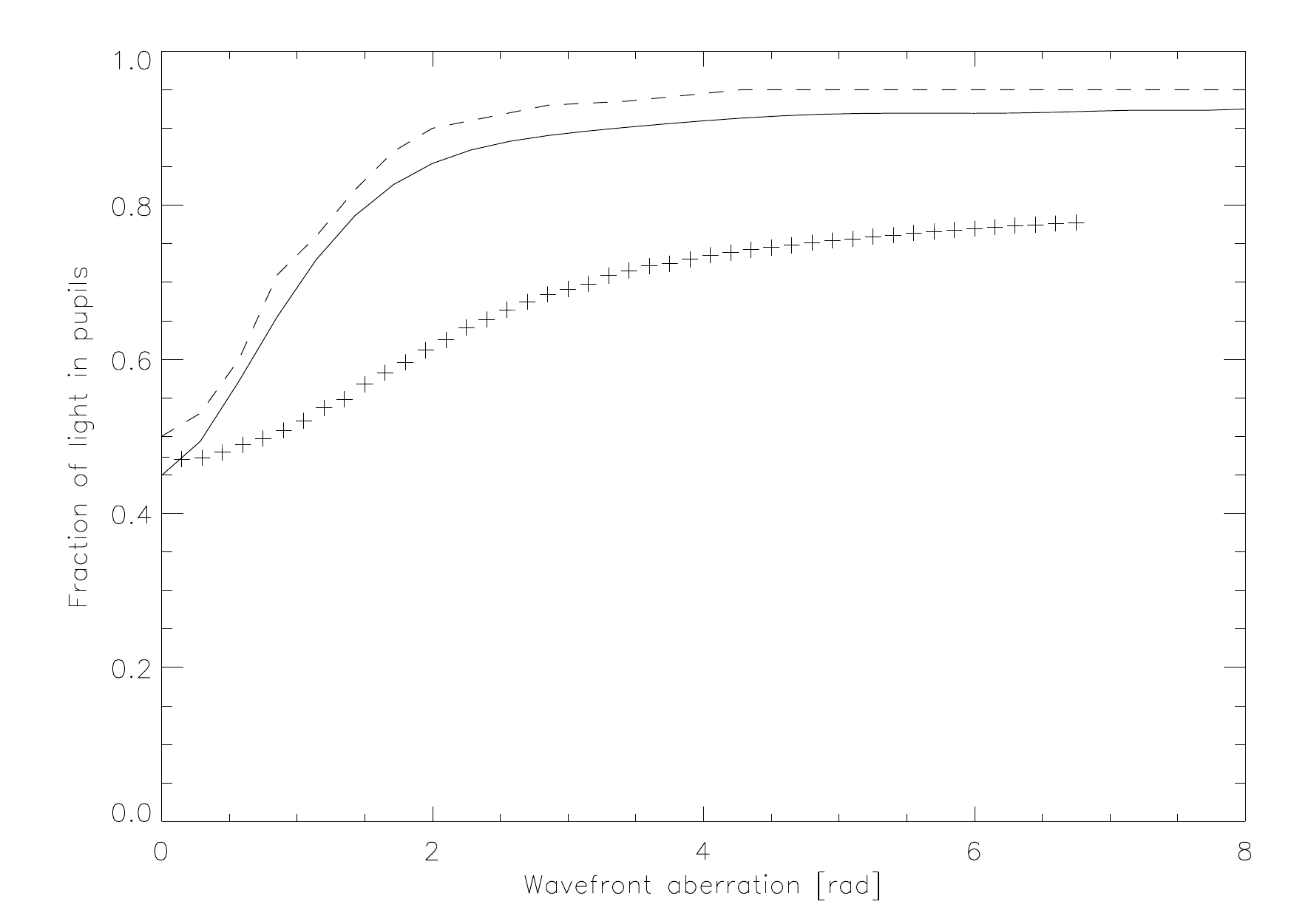}
  \caption{Light 'lost' due to the effect of the pyramid edges. The dashed
    curve is the simulated 'loss' due to diffraction only.  The solid curve
    includes the effect of light lost due to finite edge sizes. The crosses
    mark the measurements averaged over the first 5 eigen modes of the DM.}
  \label{im:f7}
\end{figure}
In Fig.\ref{im:f7} we
compare the average of the measurements of the light diffracted from the edges
(crosses) versus the amplitude of the mirror eigen modes 0-4 applied to the
DM with predictions of simulations (solid and dashed curves). 
The dashed curve shows the effect of diffraction only, whereas the solid curve
includes the effect of finite pyramid edges also. We measured the total amount of light on
the detector and the amount inside the pupils. For a (nearly) perfect
wavefront the measurement matches the simulation.  Of course it is not
possible to deduce from this measurement the ultimate cause of the light
ending up outside the pupil images. However, it is clear that apart from
the regime of very good wavefronts, we suffer additional losses that cannot be
explained by diffraction alone. From microscopic measurements we know that the
finite edge of the prism is at least in one direction more than 20\,$\mu$m
across.  A simple calculation assuming all light falling on an edge of this
size to be scattered out of the pupils, in addition to diffraction, yields the
solid curve in Fig.\ref{im:f7}. It appears that scattering on other optical
elements is important for our system for moderately well to badly-corrected
wavefronts.
\subsection{Centering on the Pyramid Tip}\label{subsec:Center}
\begin{figure}[h!]
\centering
  \includegraphics[width=8cm]{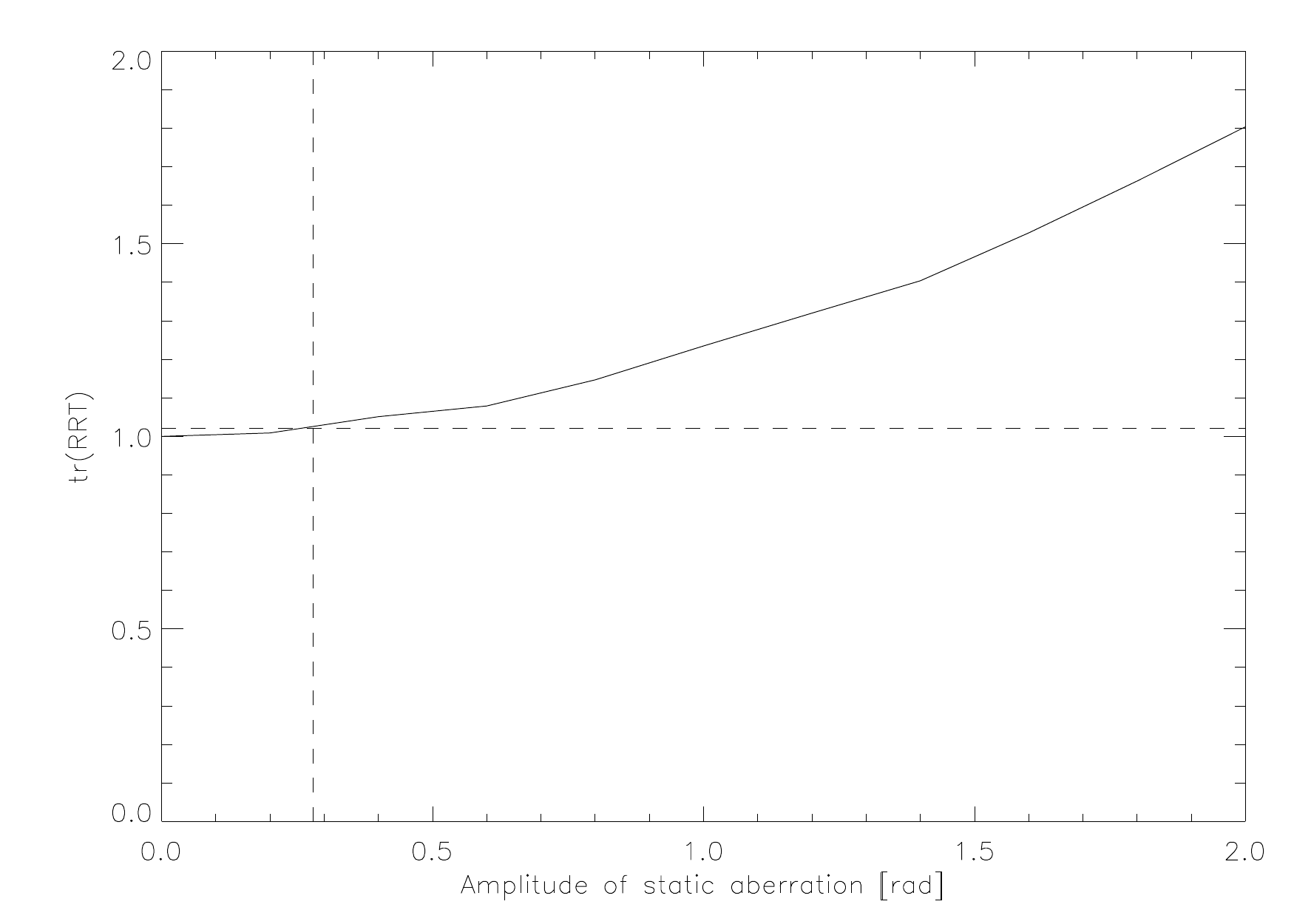}
\includegraphics[width=8cm]{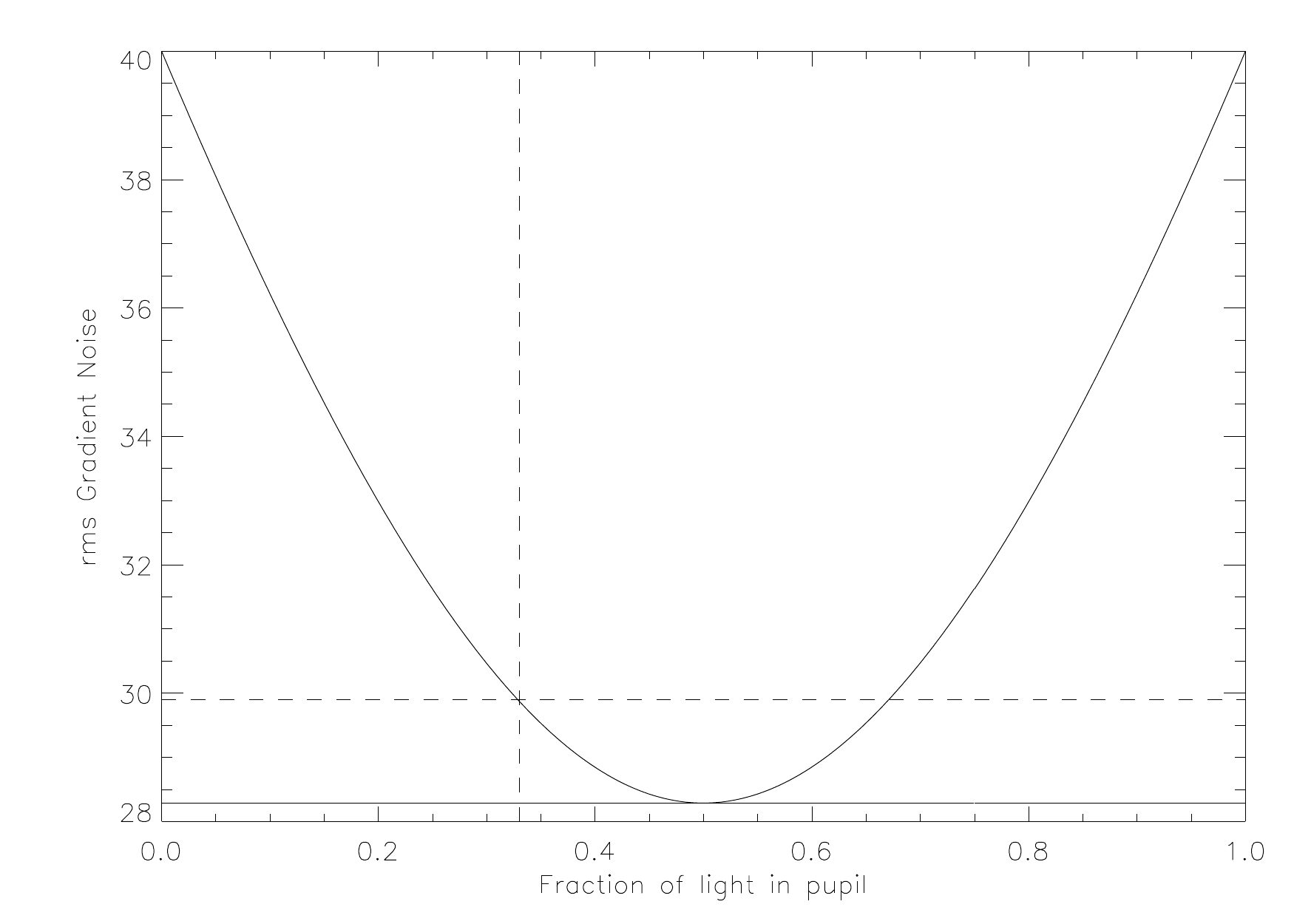}
  \caption{Left: dependence of $ tr(RR^T)$ on the amplitude of static aberrations.
    The solid line shows the dependence averaged over different modes, and the
    dashed lines mark the value for our application. In the case of a
    displacement of the dewar 0.1 $rad$ aberration correspond to $\approx 14.2
    \mu$m displacement. \newline
Right: effect of a displacement of PYRAMIR on the gradients. The
    figure shows the noise dependence of the gradients on the fraction of
    light in one of the pupils for a simplified model of a roof prism. The
    dashed line marks the momentary working point of PYRAMIR.}
  \label{im:f8}
\end{figure}
The centering of the beam on the pyramid tip is important in order to evenly
distribute the light between the four pupil images.  In the case of PYRAMIR,
the beam is centered onto the pyramid tip during calibration via the motorized
xyz-stages.  Due to the finite step size of these stages, the beam is not
perfectly centered on the tip of the pyramid during HO calibration. When the
software ignores this fact, as was the case during our early runs, the beam
will have the same decenter during CL operation.  This decenter was measured
to be small in one direction but about 40 $\mu$m in the other. This results in
two effects reducing the performance:
\begin{enumerate}
\item The reconstruction is of reduced quality
\item The noise of the gradients is increased.
\end{enumerate}
Both effects enter the measurement error 
\begin{equation}\sigma^2_{meas}=tr(RR^T)\sigma^2_{grad},\end{equation}
where $R$ is the reconstruction matrix and $\sigma^2_{grad}$ is the error of the
gradients. 
The effect on this matrix is shown in Fig.\ref{im:f8}, left panel. The effect
on the noise of the gradients is shown in the right panel of Fig.\ref{im:f8}. 
Here we show the
increase in noise per gradient versus the fraction of light in two neighboring
pupils for TT-motion. This was calculated here by using a simplified model (a
roof prism). In our case the increase in noise is 0.11. Both effects together
reduce the limiting magnitude by 0.13 mag.
\subsection{Pupil Illumination Flatness}\label{subsec:Fiber}
In the case of a PWFS, the wavefront is measured in the focal plane (not pupil
plane like the SHS). Due to this fact it is important that the illumination of
the pupil is as uniform as possible. The problem that can arise here results
from the fact that the output of a (single-mode) fiber, as it is frequently
used in AO systems as a calibration light source, has a Gaussian intensity
profile. If, for instance, the telescope has an F/10 output beam, the pupil
image has a diameter of 0.1 m, the calibration fiber has a numerical aperture
of 0.2, and the FWHM of the beam in the fiber has a diameter of 5 $\mu$m, then
at the pupil plane the FWHM of the beam is 0.1 m. 
\begin{figure}[ht!]
\centering
  \includegraphics[width=8cm]{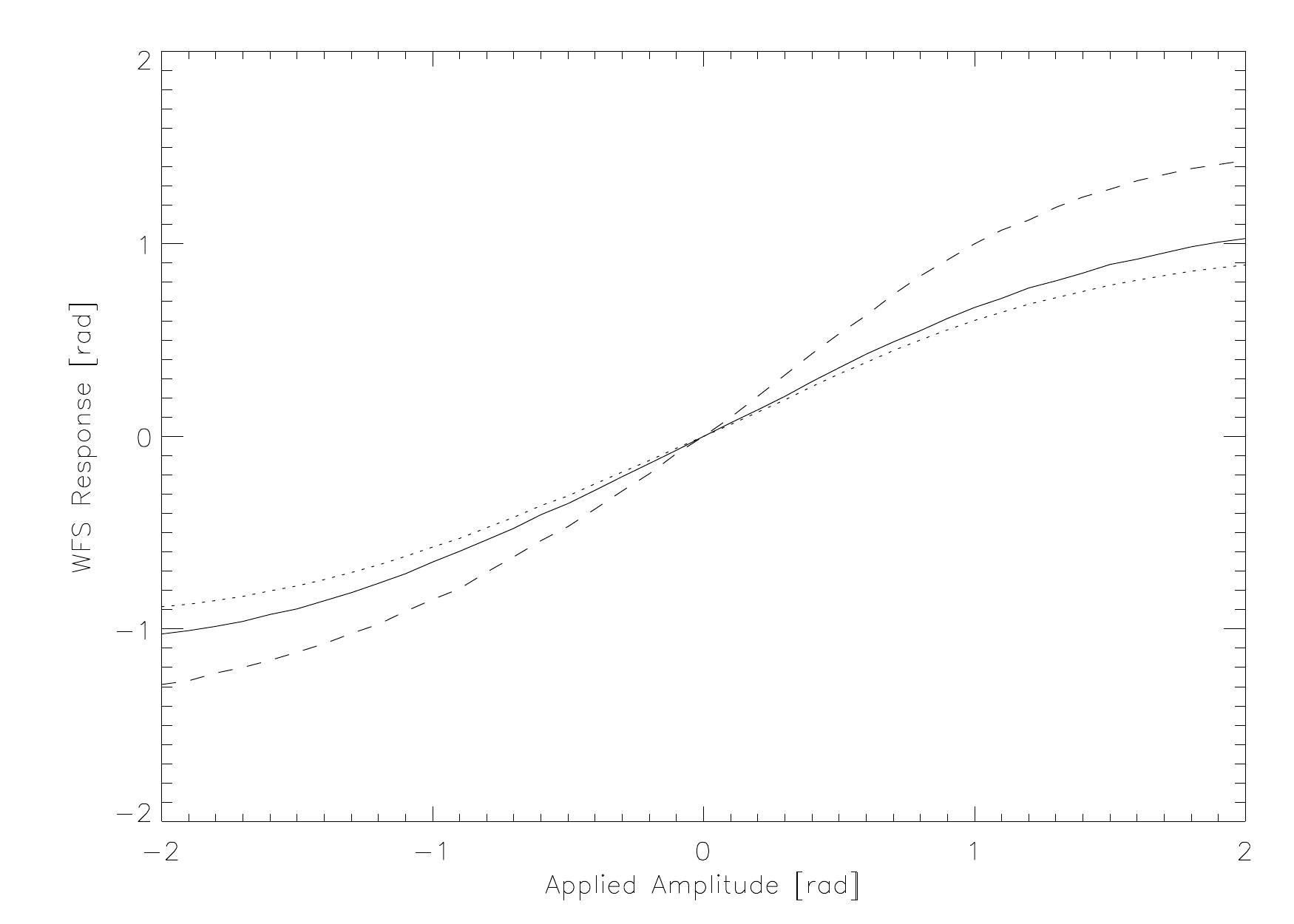}
 \includegraphics[width=8cm]{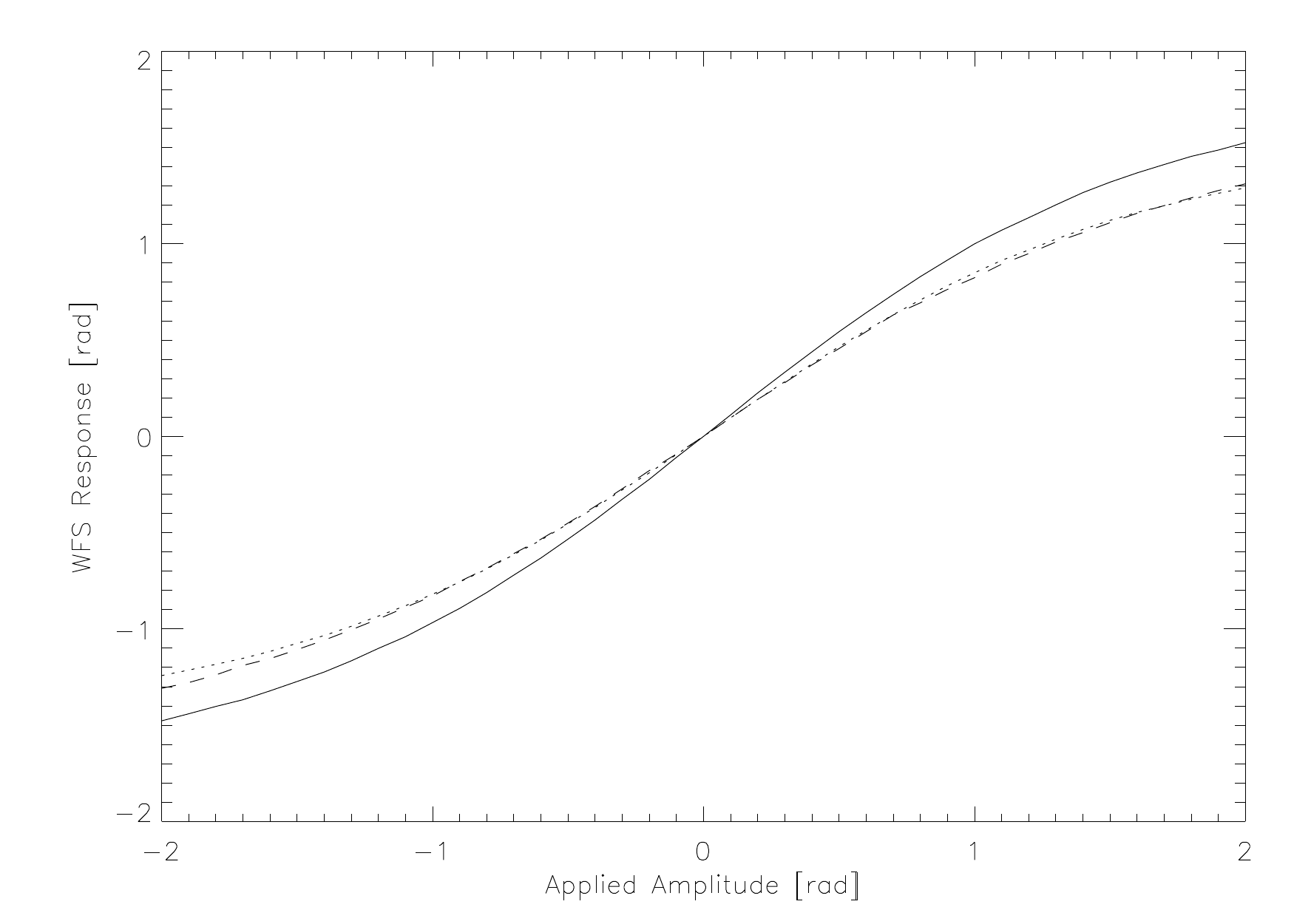}
  \caption{Left: effect of different illuminations during calibration. The
    figure shows the linear regimes averaged over the first 40 modes after changing the fiber between
    calibration and measurement. During measurement the pupil was illuminated
    by fiber 3. The dashed line corresponds to the response with fiber 3, the
    solid line to fiber 2 and the dotted line to fiber 1. The sensitivity is
    reduced in the case of a calibration with fiber 1 and 2. Right: response of the system with an extended target as the light source
    (fiber 2). An average over the first 40 modes is taken. The response of the system is almost independent of the fiber
    used for calibration.}
  \label{im:f10}
\end{figure}

This means we have an
intensity drop of 50\% from center to edge.  Simulations show that calibrating
under these conditions and then changing to a star, that, disregarding
scintillation, has a flat intensity distribution will reduce the linear regime
of the sensor (that is already small) \cite{PeterSPIE}.  For an illumination
with a FWHM of $\sigma$ the signal S on the pyramid tip will be:
\begin{equation}
  S\propto\int_{Pupil}exp\left(-{r^2\over{2\sigma^2}}\right)exp[i\psi(r,\phi)]exp[ikrcos(\phi)]r\,dr\,d\phi.\end{equation}
where $r$ and $\phi$ are polar coordinates, $\psi(r,\phi)$ is the phase of the
electromagnetic wave and $k$ is the wave number.
The effect is a smearing
of the frequencies in the focal plane in the radial direction.
For a flat wavefront the intensity distribution on the pyramid is calculated 
as \begin{equation}S\propto exp(-{\sigma^2k^2\over{2}}).\end{equation}
Thus the spot has the larger diameter of $\sigma$. It is easy to
see the influence on the TT-mode,
if the intensity distribution has a FWHM $\sigma$, then the spot in the focal
plane (on the pyramid tip) will have the size ${1/{\sigma}}$. It is
therefore wider than for an evenly illuminated flat wavefront. If it is for example twice as
wide then the TT-signals from the star will be overestimated by a factor of 2.\newline 
To solve this issue one can try two things: use a fiber with a higher
numerical aperture (NA) or with a larger core, i.e., a multimode (MM) fiber. A
MM fiber that produces a much more uniform illumination has a
much larger core diameter and might, therefore, be resolved by the
sensor. This is the case for PYRAMIR, and the result should have a similar
effect as modulation of the pyramid during calibration: The linear
regime becomes larger but the sensitivity drops.  A change
in the fiber between calibration and measurement should reveal this
fact.\newline
\begin{table}[h!]
  \caption{Properties of the fibers in use.}
  \label{fiber}
  \centering
  \begin{tabular}{c c c c c}
    \hline\hline
    Fiber & core diameter & NA & cut off ($\lambda$)  & FWHM in K\\ 
    1& 9.5 $\mu$m & 0.13 & 1400& 16 pixel\\ 
    2 & 50 $\mu$m & 0.22 & - & $>>$ 18 pixel\\
    3 & 4.0 $\mu$m & 0.35 & - & $>$ 18 pixel \\ \hline
  \end{tabular}\vspace{0.5cm}\newline

\end{table}
In the case of PYRAMIR we have the choice of three different
fibers with the characteristics shown in Table \ref{fiber}. The FWHM
mentioned there is the width of the illumination in the pupil plane,
not the focal plane.
The MM fiber can be resolved by PYRAMIR: the entrance
beam from the fiber has a F/10 optics, the PYRAMIR output a F/100. In
K band the fiber must therefore be smaller than 22$\mu$m in order not
to be resolved. Thus our MM fiber is resolved.\newline
In the following we will address two important questions:
\begin{enumerate}
\item What is the effect of a resolved fiber or a Gaussian illumination on the
  on-sky performance of the system?
\item How does an extended light source during the measurement effect the
  performance of the system?
\end{enumerate}

Fig.\ref{im:f10}, left panel, addresses question 1. There, the response of the system for
different illuminations during calibration is shown.  The fiber with high NA
was used as the light source during measurement because between the three fibers
it best resembles a point source. The sensitivity after a calibration with
fiber 1 or 2 is reduced with respect to the calibration with fiber 3 the
'true' point source.  This shows that an extended calibration fiber as well as
a Gaussian illumination, during calibration reduces the sensitivity of the
system.  \newline The effect of an extended object as the target during
measurement is shown in Fig.\ref{im:f10}, right panel.  

The response of the system is
slightly better for a calibration with the MM fiber but the difference is much
smaller than the one shown in Fig.\ref{im:f10}.  Therefore a calibration with
a perfect point source and a flat illumination of the pupil during calibration
will be the best choice even for the use of an extended object as the target for
the CL operation.
\subsection{Read out Noise}\label{subsec:RON}
Infrared detectors have large RON in comparison to CCDs.
Therefore in the regime of faint stars the RON is quite important. The effect
of the RON can be seen from the calculations of the photons needed for a given signal-to-noise ratio
(S/N) per subaperture in the dependence of the RON per pixel:
\begin{equation}N_{ph}={S/N}\left(0.5{S/N}+\sqrt{0.25\left({S/N}\right)^2+4RON^2}\right).\end{equation}
This influence will have severe consequences for the correction because the
reconstruction error is inversely proportional to $(S/N)^2$.  
\begin{figure}[h!]
\centering
  \includegraphics[width=12cm]{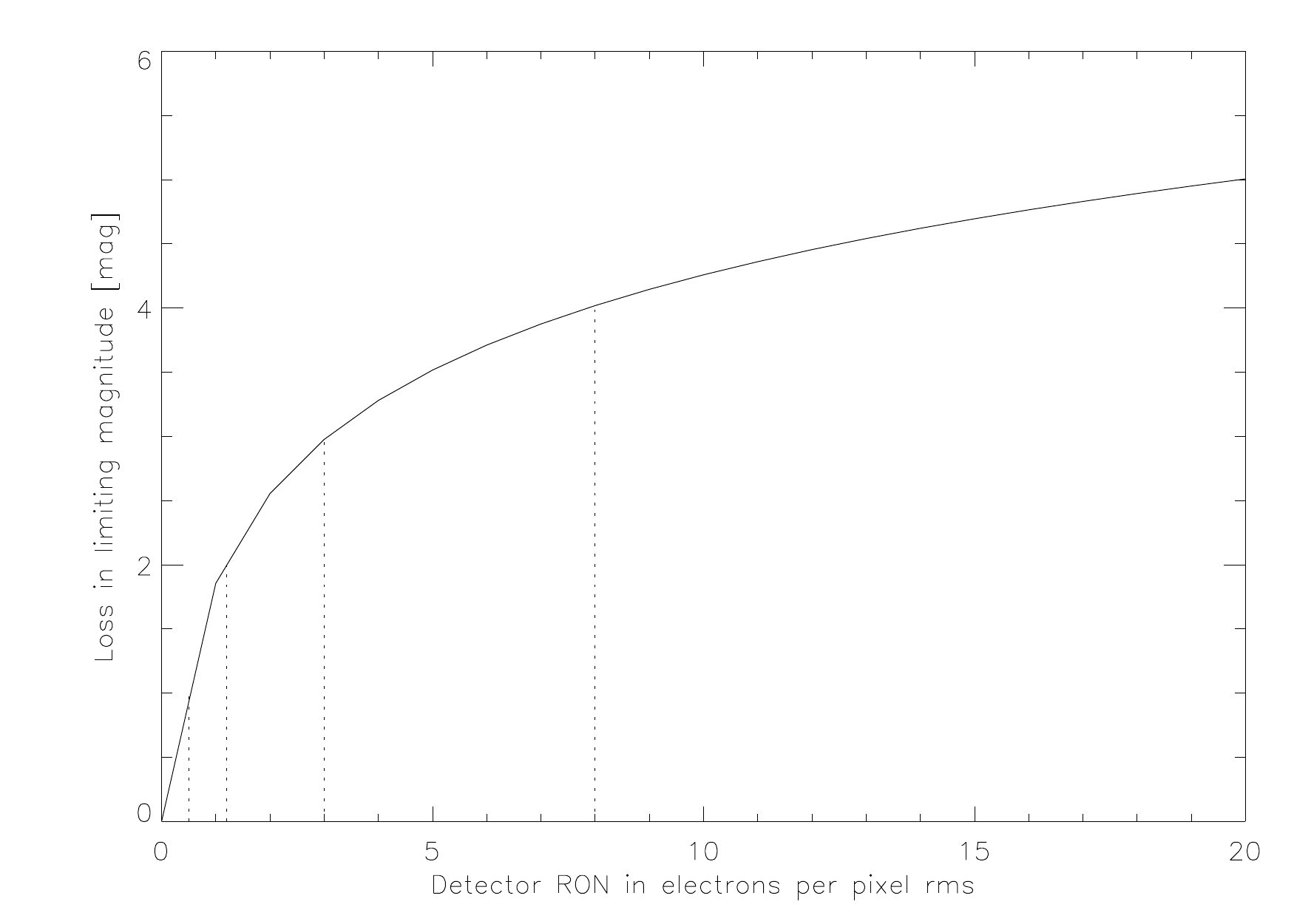}
  \caption{Loss in limiting magnitude due to RON. The dotted lines mark
    the losses in 1, 2, 3, and 4 mag. At the position of PYRAMIR with 20 $e^-$
    RON we lose 5 mag.}
  \label{im:f12}
\end{figure}
From the above
equation one can easily derive the magnitudes we lost to RON as:
\begin{equation}\Delta
  m=2.5\log\left(0.5+\sqrt{0.25+4RON^2/(S/N)^2)}\right).\end{equation}
Therefore the higher the S/N of the photon flux alone, the less important the RON. On the sky we have
shown PYRAMIR to operate well down to S/N ratios of 0.4 per
subaperture \cite{Peter2008}.
Taking this value as the limit, the loss in limiting
magnitude for a fixed Strehl ratio is shown in Fig.\ref{im:f12}. 
So for 1 $e^-$ noise per pixel only we already lose about 1.5 mag. At our level of 20 $e^-$ the
loss is 5 mag.\newline
Infrared detectors also have the
disadvantage of a slow read out compared to CCDs. Therefore they
strongly limit the bandwidth of the correction loop.
One possibility to increase the read out speed is to reduce the pixel-read
times. In  our case this increases the combined noise of detector and read out
electronics.  
\begin{figure}[h!]
\centering
  \includegraphics[width=8cm]{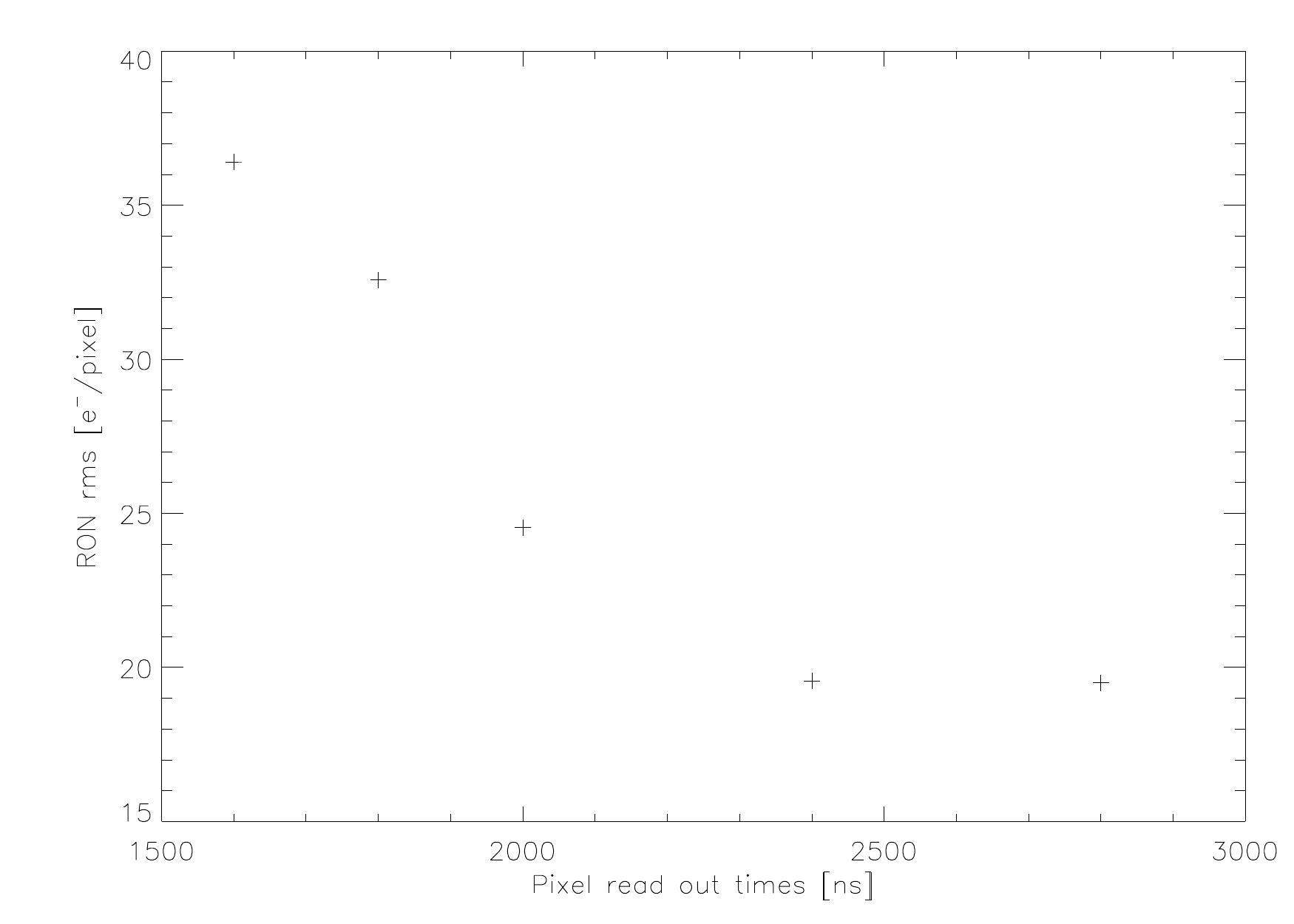}
\includegraphics[width=8cm]{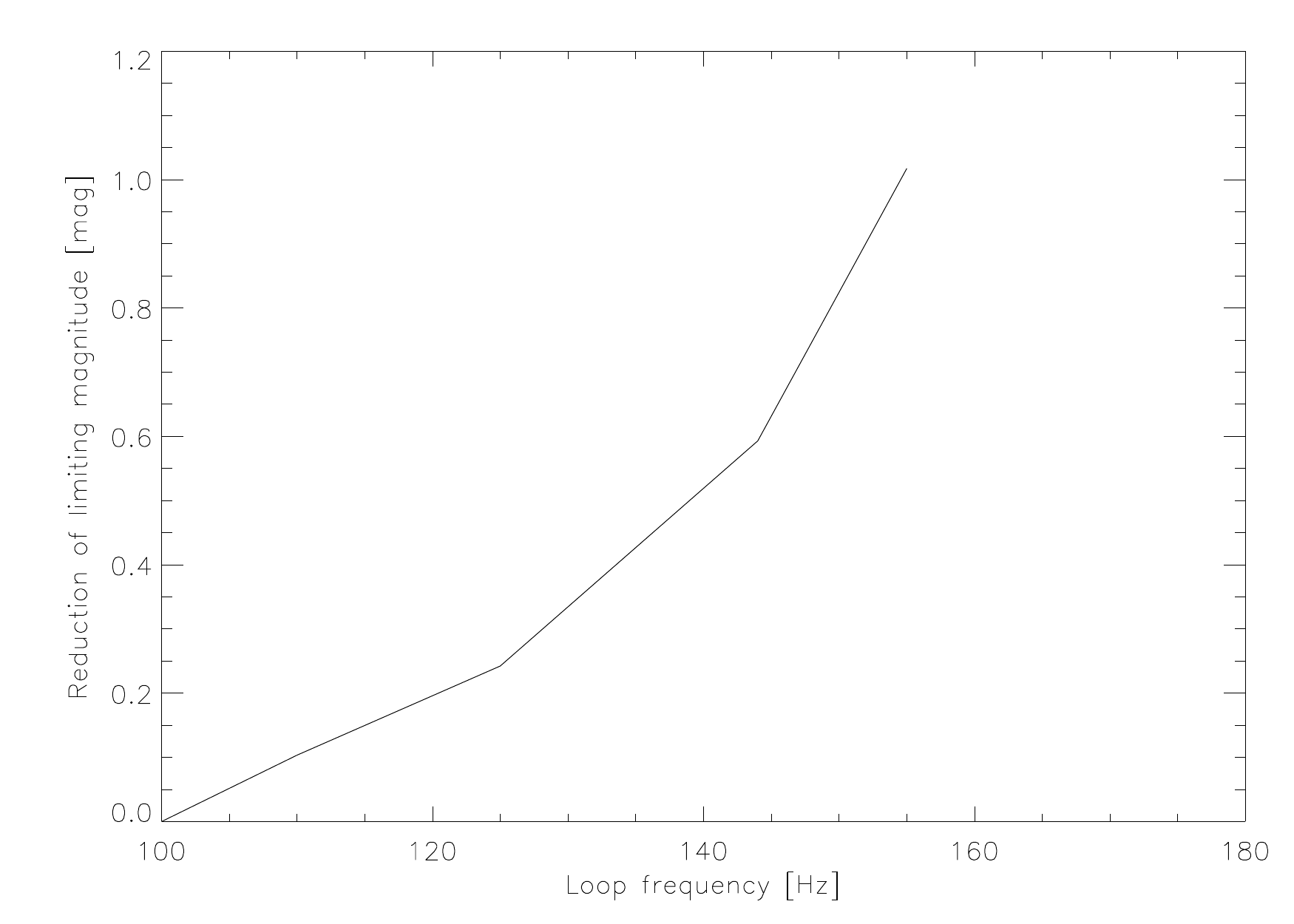}
  \caption{Left: read out noise dependence on the pixel-read times on the
    detector. Right: limiting magnitude effected by loop speed and RON compared to the
    'standard' 100 Hz loop.}
  \label{im:f13}
\end{figure}
In Fig.\ref{im:f13}, left panel,  we show the RON measured with the PYRAMIR
detector in dependence of the pixel-read times. The limit for 'slow'
read out is 20$e^-$. With decreasing pixel-read time, it
strongly rises.
If we increase the loop bandwidth by decreasing the pixel-read times the limiting
magnitude
is reduced by the combination of faster loop speed and increase in noise. Fig.\ref{im:f13}, right panel,
shows this combined effect.

\subsection{Noncommon Path Aberrations}\label{subsec:Noncomp}
Noncommon-path aberrations are unavoidably present in every AO system. Due to
the small linear regime of a PWFS it is of high importance to find a good way
to deal with them.  In the following we look into two of the three most
important properties of a wavefront sensor: the regime of linear response and
the modal cross talk in dependence of the static aberrations. \newline 
In the
linear regime the sensor will work with the best performance. The
orthogonality of the modes etc. are best in the linear regime. It should lie
symmetrically around the zero point.  The modal cross talk, i.e.  the
nonorthogonality of the modes should be as small as possible.  Thus the
mode set chosen should be well adapted to the geometry of the total system,
including the telescope and the sky. \newline Before describing the series of
measurements performed, we have to explain how a measurement is actually laid,
out.\newline Before we can start to measure we have to calibrate the system
i.e., apply modes to the DM and measure the gradient pattern on the sensor.
These measurements give us the interaction matrix ,i.e., gradients for modes.
Then the (pseudo)inverse of the interaction matrix is calculated resulting in
the reconstruction matrix.\newline
  A "measurement" is now done simply by multiplying the measured interaction
matrix with a reconstruction matrix, where the latter is possibly taken under
different circumstances, e.g., with different static aberrations applied.
\newline Further on the measurements showed that there is only a marginal
difference regarding the important properties of the sensor ,i.e., linear regime,
etc., between the use of the gradients that are measured directly and the mode
coefficients that are calculated using the reconstruction matrix of the
system. Therefore, in the following, we will exclusively use the mode
coefficients for characterization.\newline

\subsubsection{Different Mode Sets}\label{subsubsec:mose}
\begin{figure}[h!]
  \centering
 \includegraphics[width=8cm]{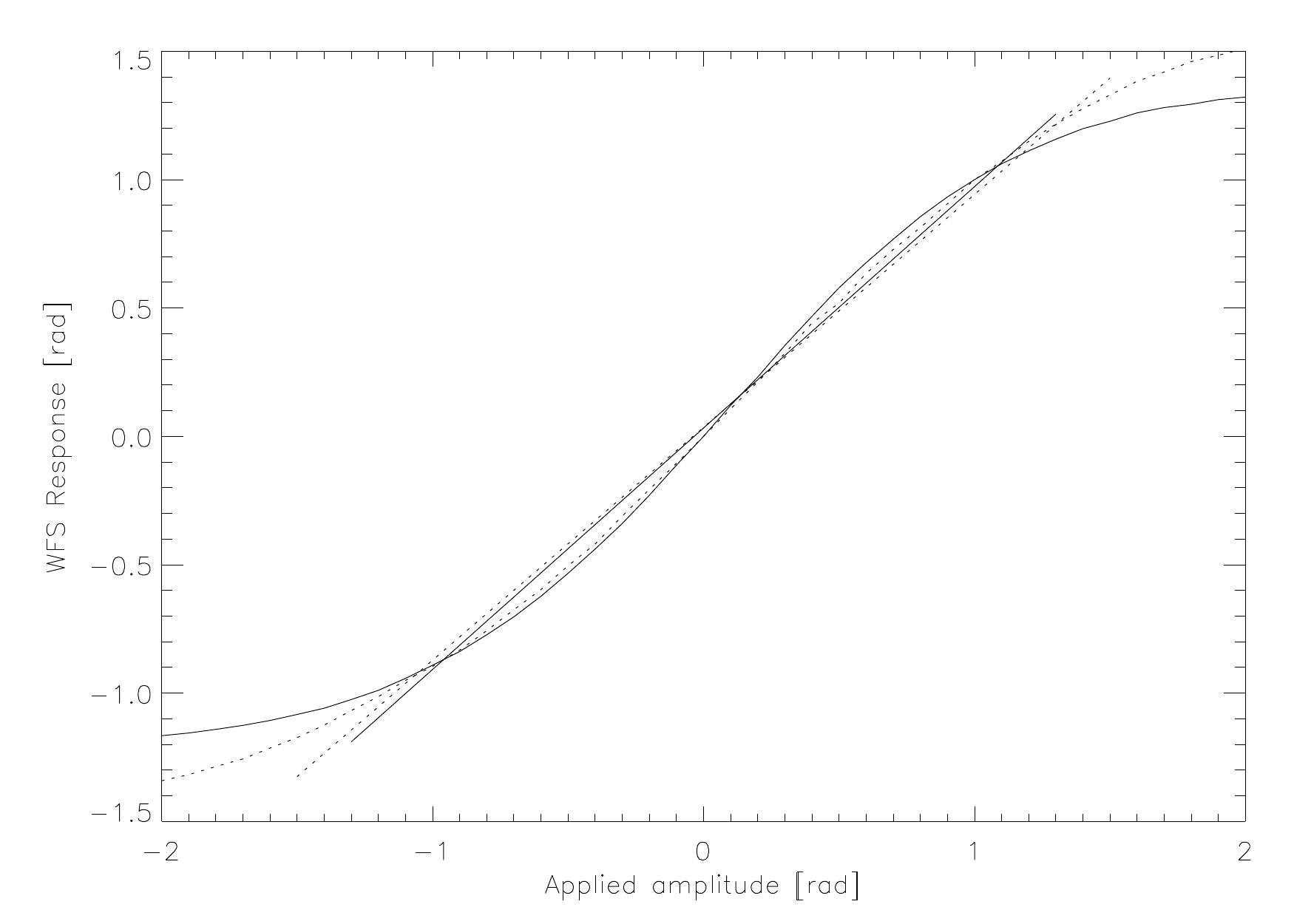}
 \includegraphics[width=8cm]{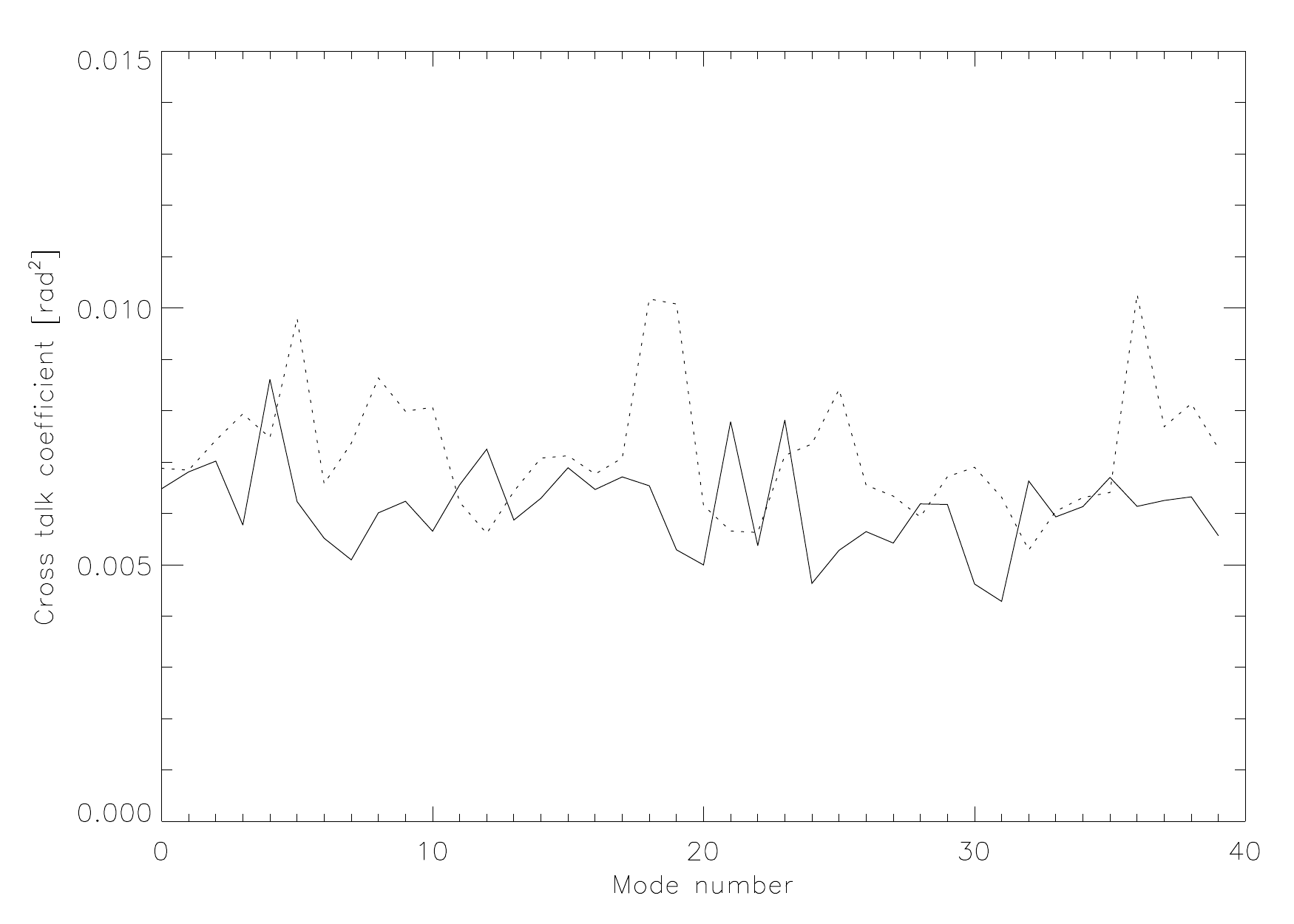}
  \caption{Left: linear regime without static aberrations for DM eigen modes ({\it solid curve})
    and KL modes ({\it dotted curve}). Shown is the linear regime averaged over 40 modes.
    The linear fit is done with a $\chi^2$ of 0.1. The linear regime of the KL
    modes is 15\% larger than that of the eigen modes. Right: modal cross talk averaged over all modes and amplitudes for each
    applied mode from -2 to +2 $rad$. The dotted line belongs to the KL
    modes, while the solid line belongs to the eigen modes of the DM.}
  \label{im:f15}
\end{figure}

In order to get a preselection of the most appropriate set of modes for the
sensor, we performed an extensive series of measurements \cite{Dor2006}. Three
different sets of modes were under examination in simulation: normalized
Karhunen-Loeve (KL) modes, eigen modes of the DM, and eigen modes of the
PYRAMIR system. The correction of atmospheric aberrations with these modes
was examined in a simulation. This simulation showed that the PYRAMIR
eigen modes are a bad choice for closed-loop operation. The reason is that one
needs a high number of these modes to reconstruct even low-order atmospheric
modes. The other two mode sets were similar in their performance. Due to the
fact that the eigen modes of the DM yield lower condition numbers, we planned
to use those for CL correction. Nevertheless the effect of static aberrations
and the linear regime of the sensor was measured for both mode sets, the DM
eigen modes and KL modes.  The amplitude of calibration varied between -2
 and +2 $rad$ (in exceptional cases from -4 to +4 $rad$) in steps of 0.1
$rad$ in K band. Several static aberrations were applied with amplitudes
ranging from 0 to 2 $rad$ rms in steps of 0.2 $rad$. From this
characterization we will find the best mode set to characterize the sensor's
dependence on static aberrations.\newline Here we compare the response of the
system for the two different mode sets.  Both sets are normalized to a rms of
1 $rad$ in K-band of the wavefront. The criteria for comparison are linear
regime and modal cross talk. The linear regime is the regime in that we can
perform a linear fit to the data with a $\chi^2$ of 0.1 or better. The modal
cross talk is the residual rms of the wavefront after subtraction of the ideal
response. Fig.\ref{im:f15}, left panel, shows the linear regime
averaged over 40 modes for both sets.  \newline 
Surprisingly, the KL functions have a larger linear range
than the eigen modes. Additionally averaged over the whole range of
calibration amplitudes from -2  to +2 $rad$ both mode sets have the same
modal cross talk (see Fig.\ref{im:f15}, right panel).\newline However, there are some modes in the
set of KL functions that naturally yield a high modal cross talk with others.

The future task will be to optimize the mode set further in order to minimize
the total modal cross talk of the entire set of basis functions and still have
an acceptable performance on sky.\newline In the following, for laboratory
purposes, we used the eigen modes unless stated otherwise.  The reason is that
the smaller linear regime helps to see the effects of static aberrations
better.
\subsubsection{The Linear Regime and Modal Cross Talk under the Influence of
  Static Aberrations}\label{subsubsec:linreg}
In this subsection we will discuss the measurements of the system's behavior
under static aberrations.  In order to keep things simple and still see the
general properties, static aberrations will be represented by one single mode
with variable strength.\newline There are differences in the effect of the
statics depending on if they are applied during calibration only, during measurement
only, or both.\newline 1. Static aberration during calibration only: this case
is rather of academic interest.\newline In this case one would expect little
effect on the modes without statics during the measurement as long as the
static mode is within the linear regime. If it is beyond the linear range
there will be enhanced modal cross talk. The exception will be the mode with a
static part. If the total amplitude of this mode exceeds the linear range, then
the response curve during measurement should become less steep because during
calibration the applied modes will be underestimated.\newline 2. Static
aberration during measurement only: this will happen if one wants to have the
best possible calibration and then a perfect science image moving the
noncommon-path aberrations entirely into the sensor path during
observation.\newline Here one would expect little effect on the modes without
static pattern unless the mode is beyond the linear range. For the modes with
static pattern the center of the response curve will be shifted. The amount of
the shift should be just the amplitude of the static mode. Additionally one
would expect the modal cross talk to increase.\newline 3. Same static
aberrations during calibration and measurement: this will be the case if we
flatten the wavefront on the science detector and start calibration
afterward.\newline Here we will expect little effect on the modes without
static pattern as long as it stays inside the linear range. The modes with
static parts will differ only slightly in the direction of the calibration
amplitude.  When the signal has amplitude either 0 or equal to the calibration
amplitude, the response for the mode will be equal to the optimum response.
In between the difference will not be high. In the other direction they will
diverge from the optimum response curve. This divergence will depend on the
strength of the static mode and the calibration amplitude. If those have, for
instance, opposite signs then we will calibrate in the linear range and end up
with the case of statics during measurement only. When they have the same
sign, we will have a similar behavior as statics during calibration only and,
thus, an underestimation of the amplitude in this direction.\newline 
\begin{figure}[h!]
\centering
  \includegraphics[width=8cm]{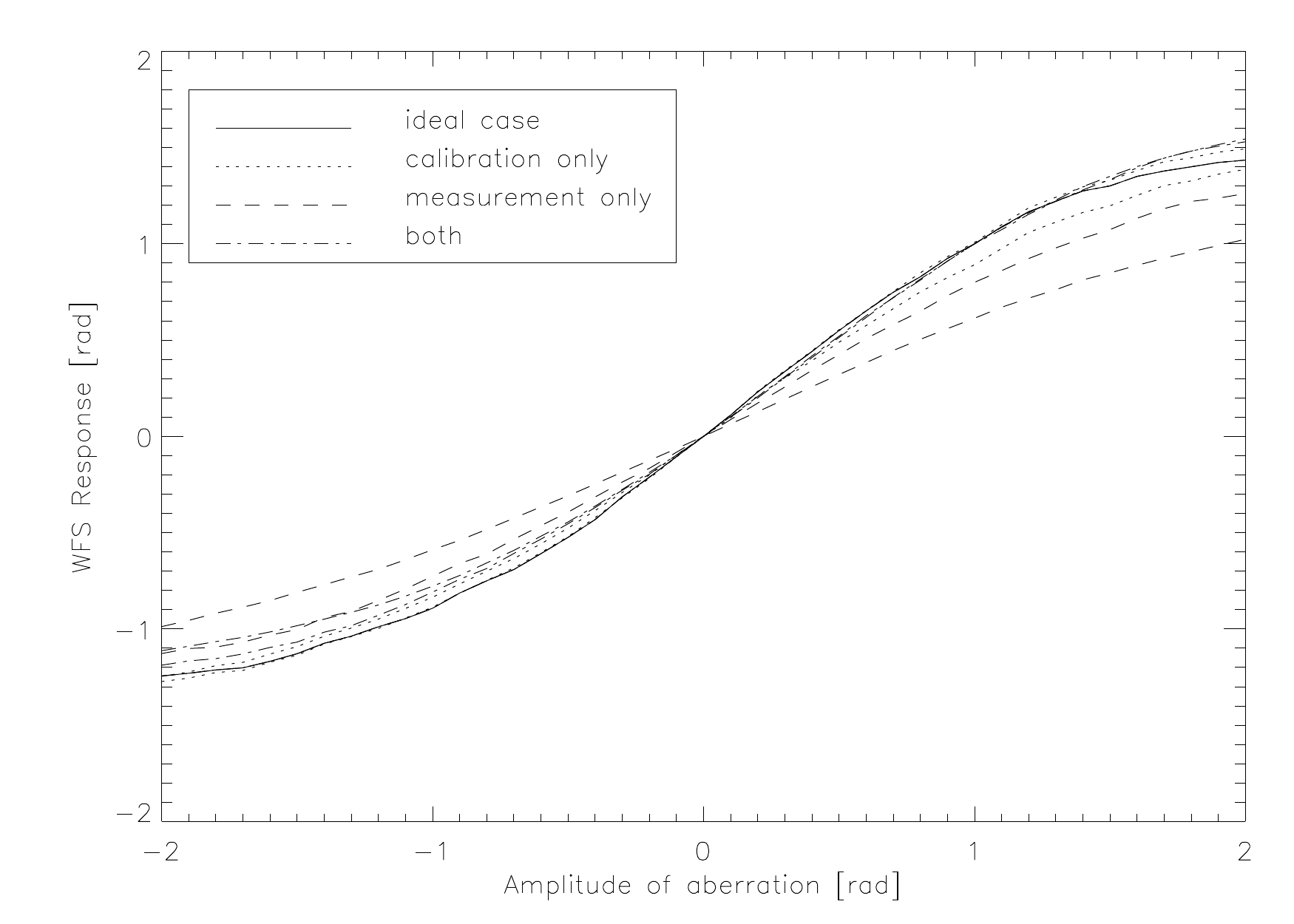}
\includegraphics[width=8cm]{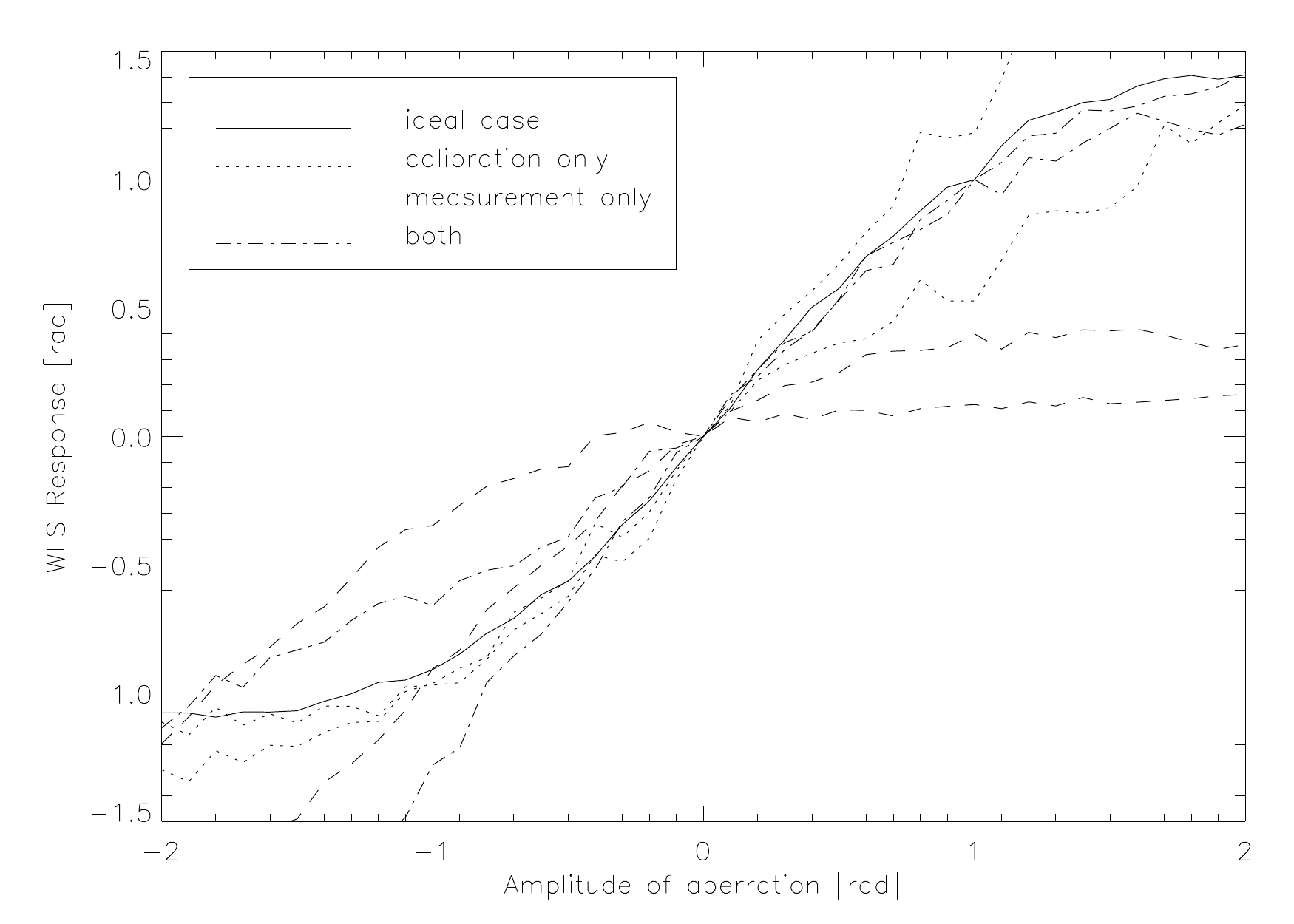}
  \caption{Left: linear regime of PYRAMIR in K. The linear regime is averaged over
    all modes. The static aberration amplitude is 1 and 2 $rad$. Shown are the
    ideal case without aberration ({\it solid curve}), statics during calibration only
    ({\it dotted curve}), statics during measurement only ({\it dashed curve}) and statics during the
    whole process ({\it dash-dotted curve}). The divergence from the ideal case goes with
    the strength of the static aberration. Only the dashed and dotted curves
    in the case of 2 $rad$ static aberration differ significantly from the
    ideal case. Right: response of the mode with static aberration. The curve
    styles are the same as the left panel. The difference between the
    cases is clearly seen. Here again the strength of divergence from the
    ideal case goes with the strength of the statics. The case closest to the
    ideal one is that with static aberrations during the entire procedure.}
  \label{im:f17}
\end{figure}
After this
short discussion of the expectations we will compare the theoretical
considerations with the actual measurements. Fig.\ref{im:f17}, left panel, shows the
averaged linear regime for the three cases in comparison with the ideal case.
The calibration amplitude was 1 $rad$, and the static aberrations are applied with
amplitude 0,1,or 2 $rad$. Only in the case of the strongest static aberration
during measurement only the response differs significantly from the other
curves. This finding matches the expectations quite well. The divergence
arises from the mode with static part, as we will see in the following. 
\begin{figure}[h!]
\centering
  \includegraphics[width=12cm]{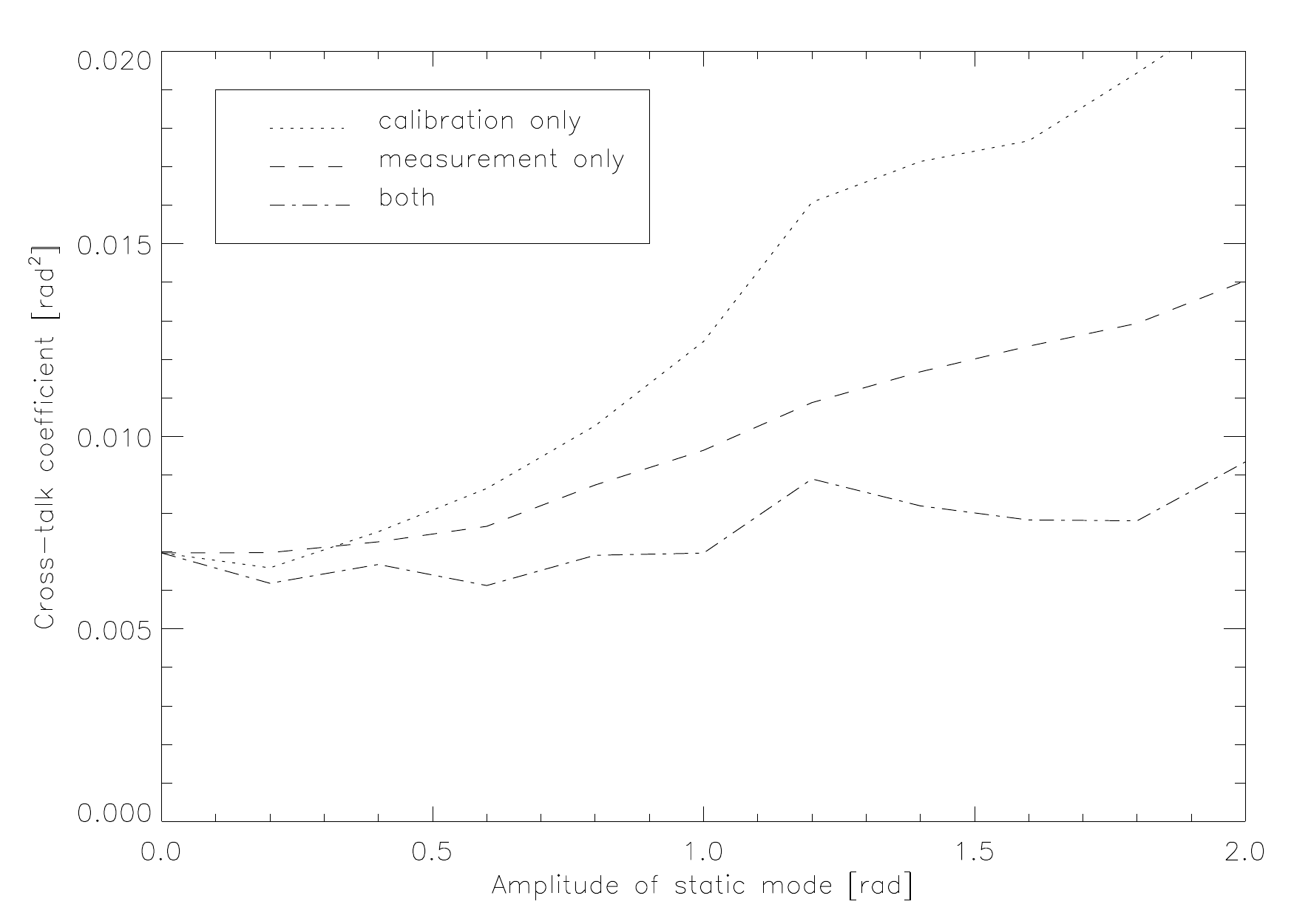}
  \caption{Modal cross talk coefficient averaged for all applied amplitudes from -2 to +2
    $rad$ vs amplitude of static aberration. The higher the amplitude, the
    stronger the modal cross talk In the case of the same static aberration
    during calibration and measurement the modal cross talk is rising only
    very slightly. The line styles are as in Fig.\ref{im:f17}.}
  \label{im:f19}
\end{figure}
 The
response of the special mode with static part is shown in Fig.\ref{im:f17}, right panel.
All curves differ from the ideal case exactly as expected. However, the
response of the system with statics applied during both calibration and
measurement is closest to the ideal case. This property should also have an
effect on the modal cross talk between the modes, as we will show
immediately.\newline

In Fig.\ref{im:f19} the averaged modal cross talk
coefficient of the
system in dependence of the amplitude of the static aberration is shown. It
rises as expected for the cases with static aberration during measurement or
calibration only. In the case of statics during both, the modal cross talk is
nearly independent of the static mode. Therefore we can conclude that in the
case of noncommon-path aberrations and the wish of a perfect science image, it
is best to move the aberrations all into the sensor path before calibrating
the system.\newline The real noncommon-path aberrations for the system were
measured to be maximum 0.4 $rad$ rms for a single mode or added quadratically
for all modes something like 0.6 $rad$ in K. Thus one is still in the linear
regime (see Fig.\ref{im:f15}) of PYRAMIR and comparing these numbers with our
result from above (see Fig.\ref{im:f17}, left panel, \ref{im:f19}) we see that we can
happily use this shape of the DM (dmBias) as starting point for calibration
and measurement. \newline An even better way to treat the noncommon-path
aberrations would be to use to different calibrations: one calibration with
positive amplitude and one with negative. Then in CL the modes with positive
and negative amplitudes will be corrected by the corresponding reconstruction.
This should reduce the measurement errors on the modes with a static part
significantly. however, up to now this did not run stably on the sky.
Therefore in the following we will always use the same statics during
calibration and measurement.
\subsection{Dependence on the Calibration Amplitude}\label{subsec:Ampl}
Here we investigate into the dependence of relative modal cross talk, aliasing
and the measurement error on the amplitude of calibration. \newline For small
calibration amplitudes the strength of the mode signal is comparable to the
noise on the detector, and we, therefore, expect the errors to decrease with
rising calibration amplitude at least within the linear regime. Outside the
border of the linear range the errors will increase again because of a rising
nonorthogonality of the modes. \newline 
\begin{figure}[h!]
\centering
  \includegraphics[width=8cm]{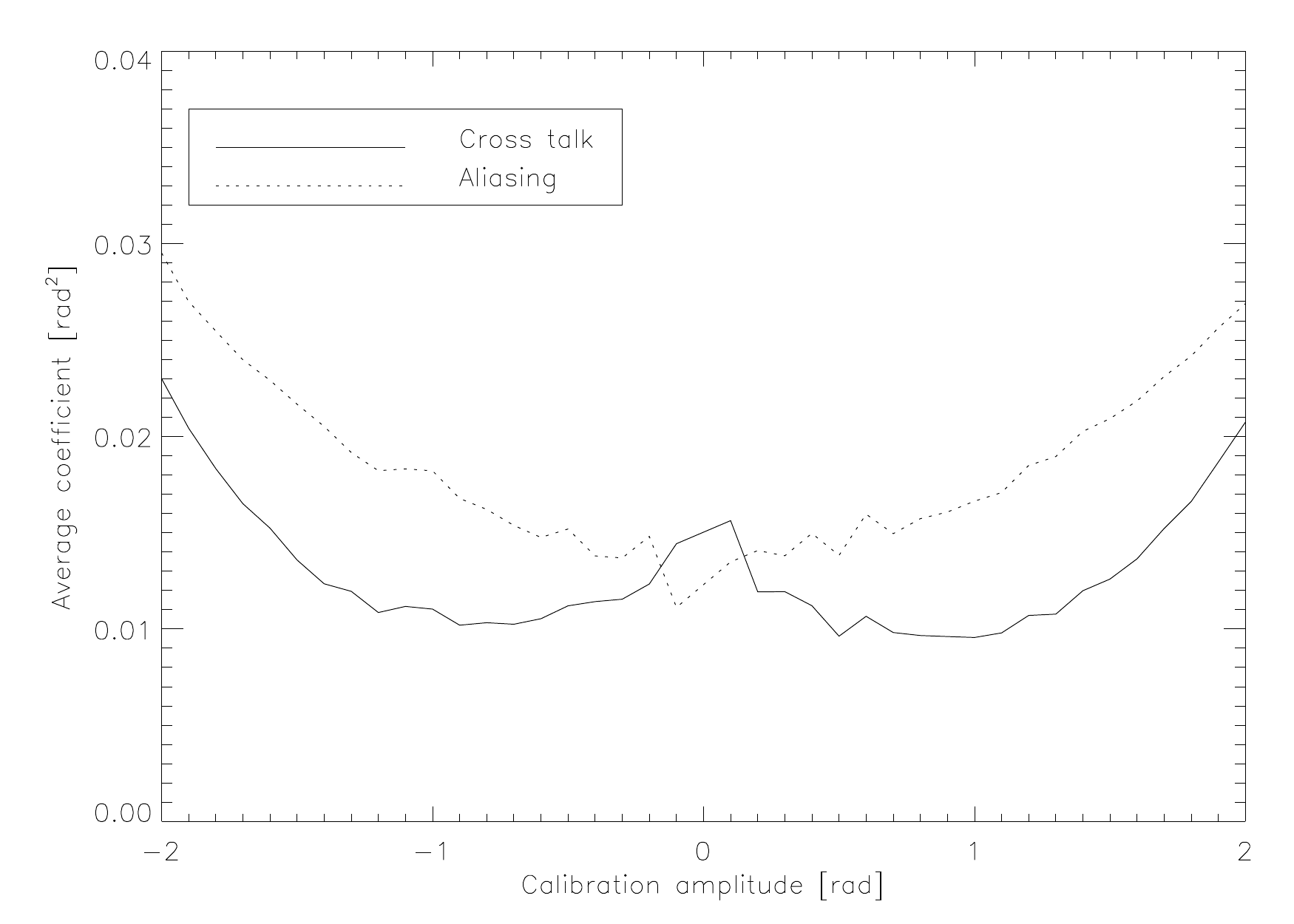}
 \includegraphics[width=8cm]{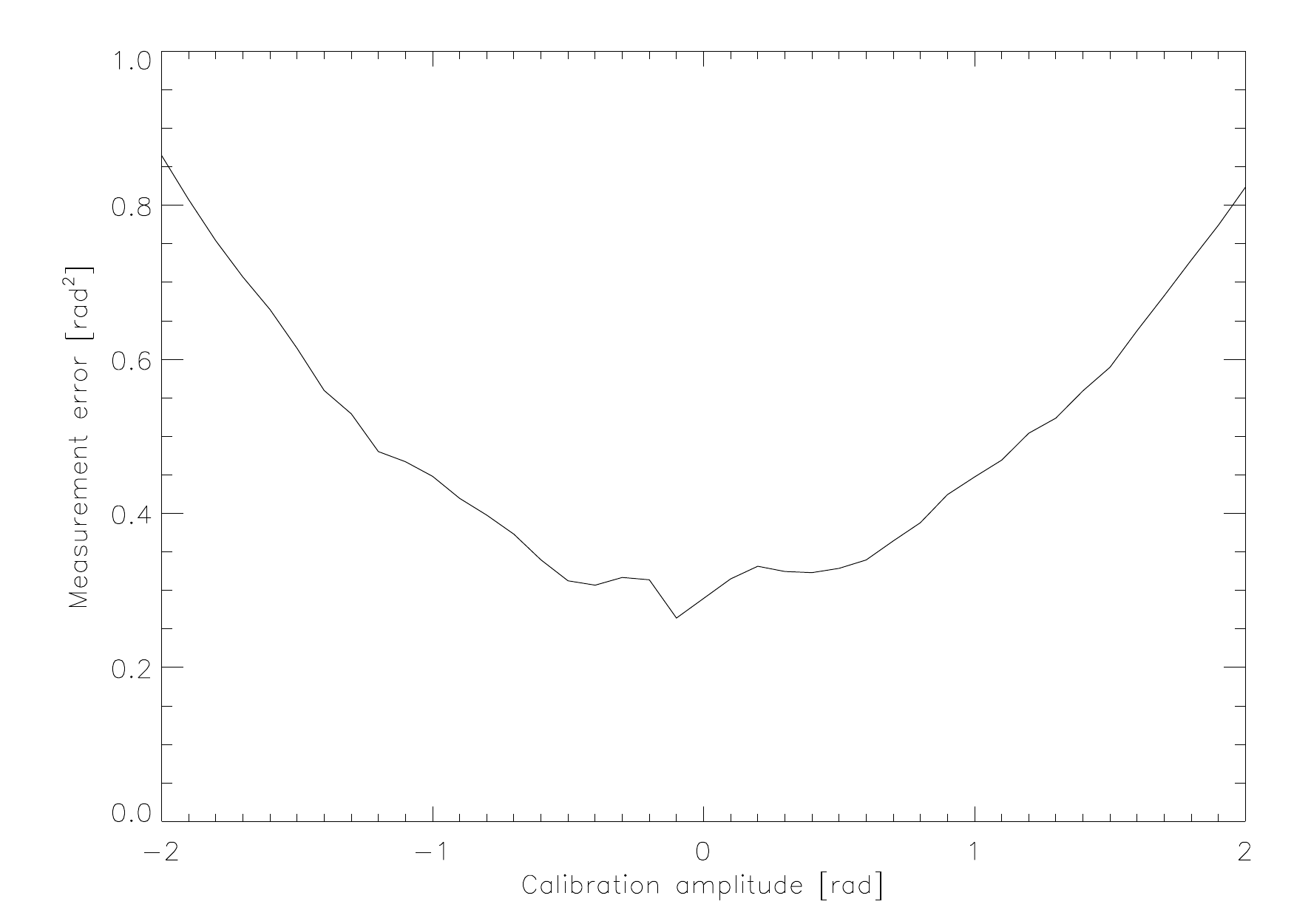}
  \caption{Left: dependence of the average modal cross talk ({\it solid line}) and aliasing ({\it dotted
    line}) coefficient for 20 calibrated modes on the amplitude of
    calibration. Right: same as in the left panel but for the measurement error in the
    case of S/N=1.}
  \label{im:f20}
\end{figure}
Fig.\ref{im:f20}, left panel, shows the dependence
of the modal cross talk and aliasing coefficients on the amplitude of calibration.
The expectations are quite well matched. The modal cross talk is high for
small calibration amplitudes and drops toward the border of the linear regime
at about 1.2 $rad$. Then it rises again. The aliasing error rises almost
parabolic with calibration amplitude but stays within 10\% difference in the
linear regime.  
The measurement error is almost constant within the regime of
-0.6 to 0.6 $rad$ and then rising more steeply as shown in Fig.\ref{im:f20}, right panel.
Therefore we conclude that a calibration amplitude of about 0.6 $rad$ will
minimize the total reconstruction error.
\begin{figure}[h!]
\centering
  \includegraphics[width=12cm]{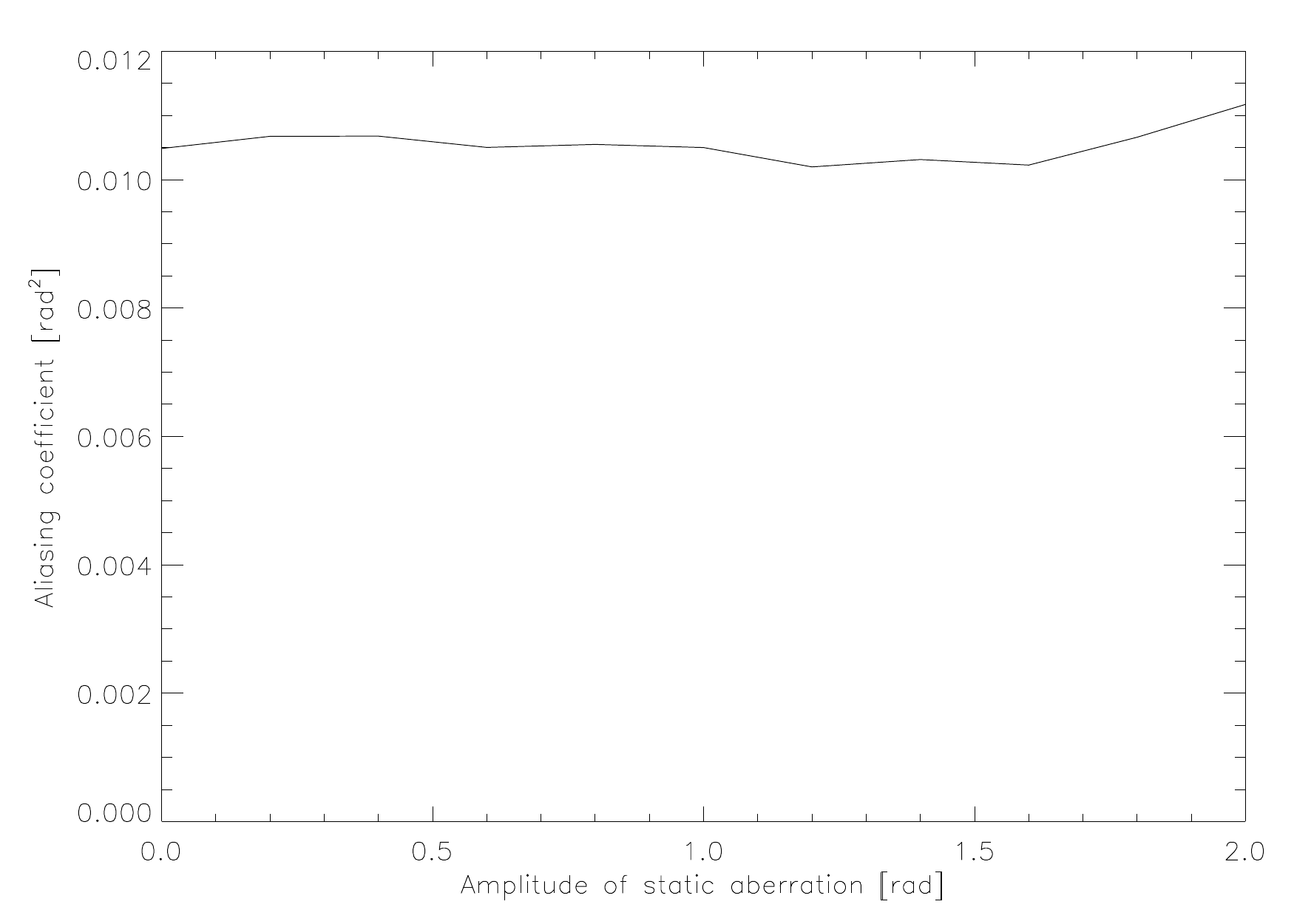}
  \caption{Dependence of the aliasing error coefficient on the strength of the static
    aberration.}
  \label{im:f22}
\end{figure}
\subsection{Dependence on the Strength of Static Aberrations}\label{subsec:Aberrstr}

We have already seen in section \ref{subsec:Center} that the measurement error
depends linearly on the strength of the static aberration. This dependence
, however, is small. In section \ref{subsubsec:linreg} and Fig.\ref{im:f20}, right panel, we
showed that there is no dependence of the averaged modal cross talk on the
statics.  The same is true for the aliasing: 
Fig.\ref{im:f22} shows the
dependence of the aliasing coefficient on static aberrations up to a strength of 2 $rad$.
The response is averaged over all numbers of calibrated modes. This dependence
will be different when the static mode is only present during closed loop.
\section{The Effect of the Errors in Closed Loop Operation}\label{sec:InflCL}
After we measured the modal cross talk of the system for different cases, one
general question arises: how much modal cross talk can we tolerate? This
question is tightly connected to the problem of how many modes are appropriate to
calibration.  We will see that from the dependence of the total residual
wavefront error is dependent on the number of calibrated modes. There are
several errors that depend on this number:
\begin{enumerate}
\item The modal cross talk between the calibrated modes $\sigma^2_{cr}$
\item Aliasing $\sigma^2_{alias}$
\item The measurement error $\sigma^2_{meas}$
\item The temporal error $\sigma^2_{temp}$
\item The fitting error $\sigma^2_{fit}$.
\end{enumerate}
Before solving the total problem we have to investigate into the nature of the
single errors. To derive these we have to take the strength of the atmospheric
modes taken from \cite{Hardy1998},
\begin{equation}\sigma^{2}_{fit}\approx0.3\left({D\over{r_0}}\right)^{5\over3}(N_m)^{-{{\sqrt{3}}\over{2}}}\label{fiterr}\end{equation}
\begin{eqnarray*}r_0&=&\,Coherence\,length\\
  D&=&Telescope\,diameter\\ N_m&=& Number\,of\,free\,parameters\,(modes)\\
\end{eqnarray*}
into account. The equation describes the residual wavefront error after a
perfect correction of the first $N_m$ modes. Thus the difference between the
results for $N_m$+1 and $N_m$ modes describes the error connected with mode
$N_m$ alone. Also we have to include the correction of the modes in CL
operation. This is the point where the temporal error comes in during this
investigation, because the band-width of the modes increases with mode
number.\newline Modal cross talk and aliasing: \newline both have quite similar
effects but not identical. Aliasing describes high-order modes we have not
calibrated, but that show up as lower-order modes thus giving erroneous
signals. Modal cross talk , however, arises between modes we know. Therefore we
get erroneous signals between all the modes we calibrated. But due to the fact
that we also measure the real mode signal we are able to reduce their
amplitudes in CL and, therefore, the strength of the error with respect to the
aliasing case. \newline The aliasing error strongly depends on the geometry of
the system. For a PWFS it is proposed to be much smaller than for the SHS (see
\cite{Ver2004}). Here we have measured the aliasing for calibrations of 2 to
39 HO modes. Due to the fact that we originally measured 40 modes we can only
use the residual 38 to 1 not calibrated modes respectively to derive the
aliasing error. 
\begin{figure}[h!]
\centering
   \includegraphics[width=12cm]{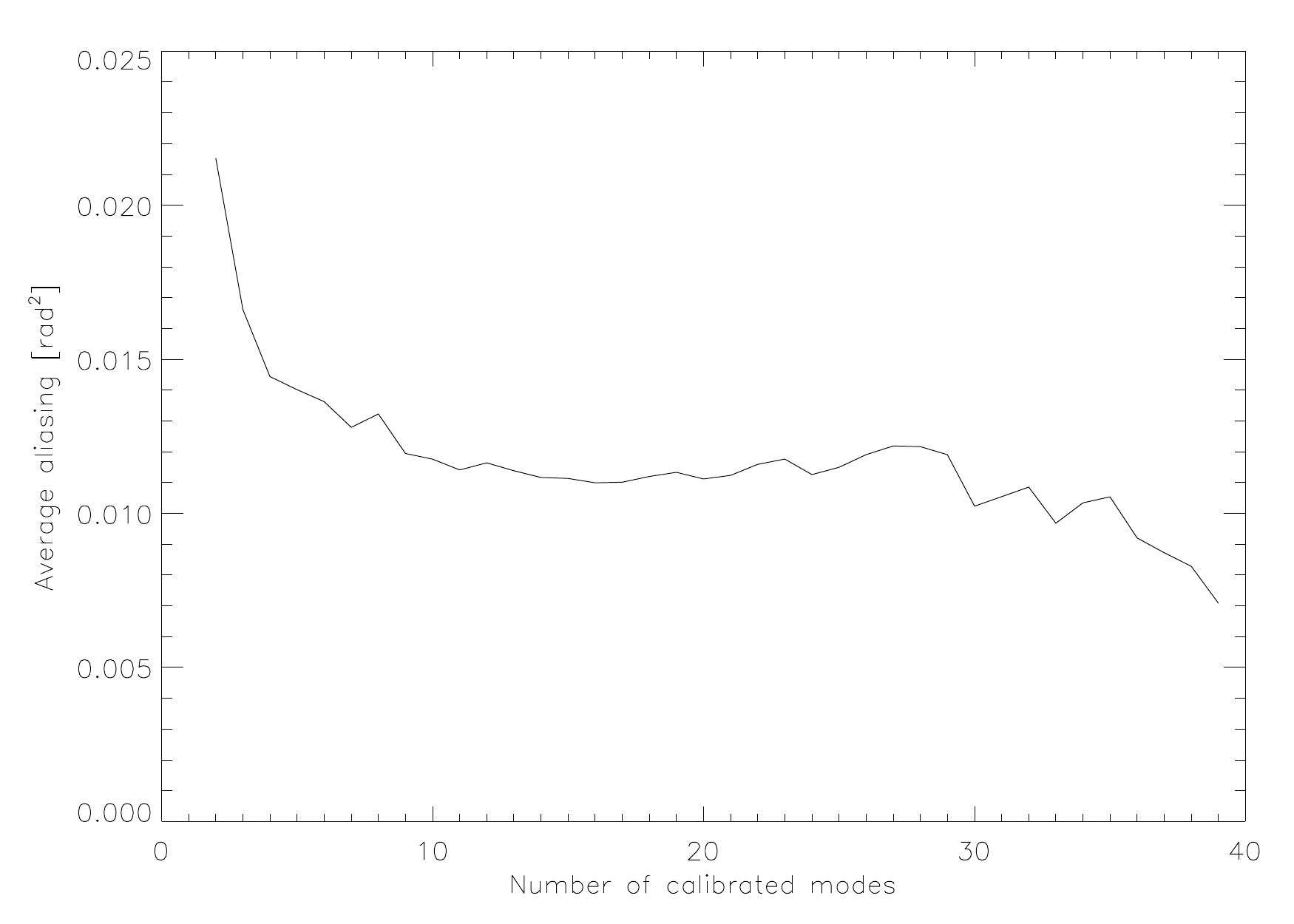}
   \caption{Development of the average aliasing coefficient with the number of
     calibrated HO modes. First it drops steeply, then stays almost constant.}
   \label{im:f23}
 \end{figure}
Fig.\ref{im:f23} shows the average aliasing error coefficient
$E^2$ dependence on the number of modes calibrated in the system. 
The coefficient decreases with 
the number of calibrated modes for up to 10 modes then it stays
constant. Still for our purposes it is sufficient
to assume it to be constant for all calibrations without making too large a mistake.
To derive the total aliasing error on sky we still have to account for the
modes not measured in the lab (No. 41 and so on). To include these we just
assumed the aliasing coefficient to be constant for all modes. We then only
had to weight every mode with its strength on sky.  The total
error due to aliasing for $N$ calibrated modes is then given as,
\begin{equation}\sigma_{al}^2=\sum_{i=N+1}<a^{\perp T}_i R^T
  Ra^{\perp}_i>=NE^2 \sum_{i=N+1}
  a^{\perp 2}_i\propto
  0.3\left({D\over{r_0}}\right)^{5\over3}E^2N^{{2-\sqrt{3}}\over{2}}\end{equation}
Here $a_i^{\perp}$ denotes the mode vector orthogonal to the set of calibrated modes and $R$ the reconstruction matrix.
The error rising slowly with the number of calibrated modes. 
We used the formula (eq. \ref{fiterr}) for the fitting error to account for the total strength of
each mode. \newline
The modal cross talk of the modes does not only depend on the systems geometry but
also on the mode set chosen for calibration. Of course the set should be
close to the KL modes. The reason is that the strength of these modes falls
off quickly as $\approx N_m^{-(\sqrt{3}+2)/2}$ with mode number and one does not want to lose the advantage of a
good correction with a relatively low number of modes. For the higher order
modes this is not so crucial because the fall off becomes almost flat. 
But still we have a bit of freedom here to reduce the total amount of
modal cross talk for a given system by the optimization of the mode set.\newline
\begin{figure}[h!]
\centering
   \includegraphics[width=8cm]{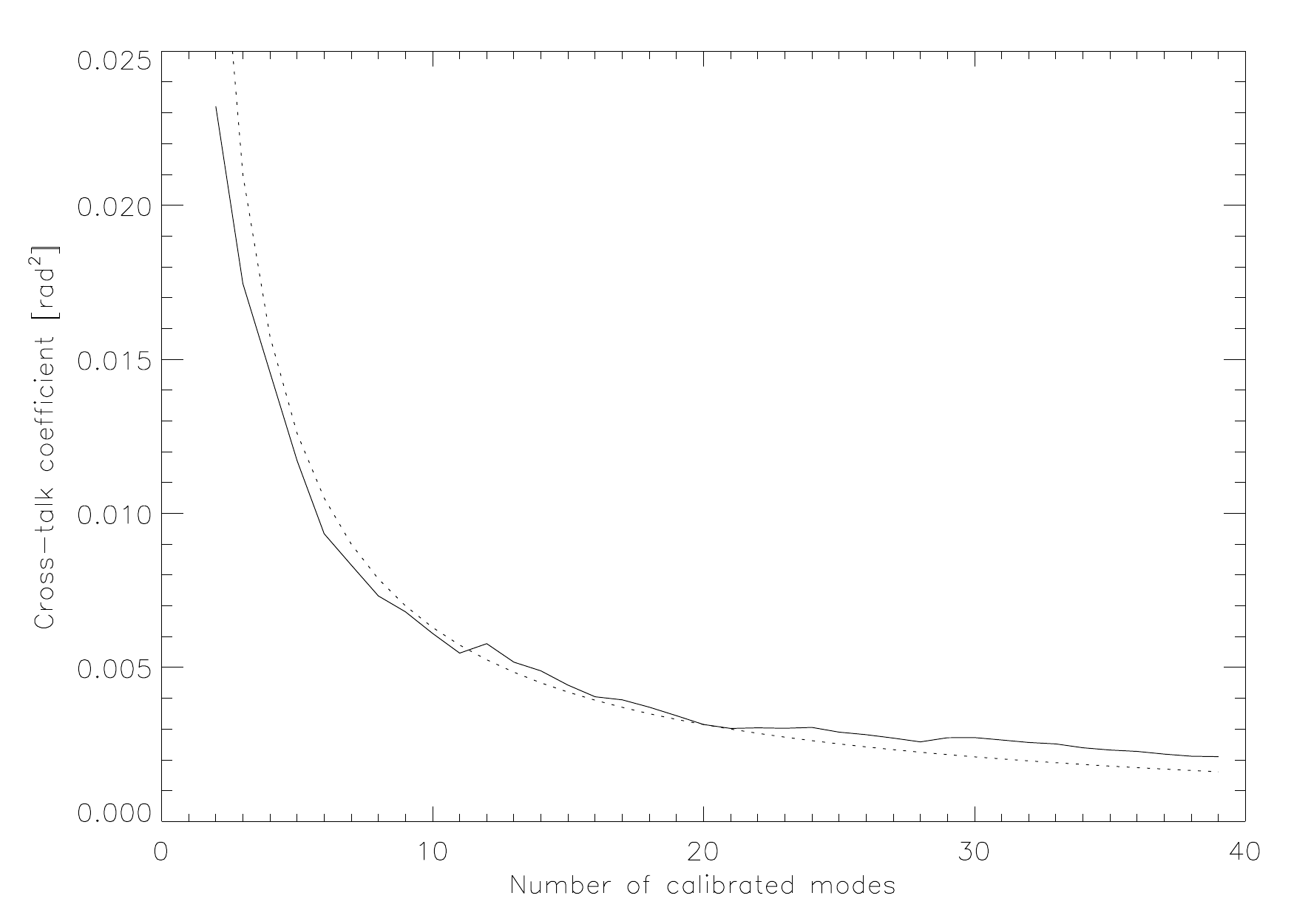}
\includegraphics[width=8cm]{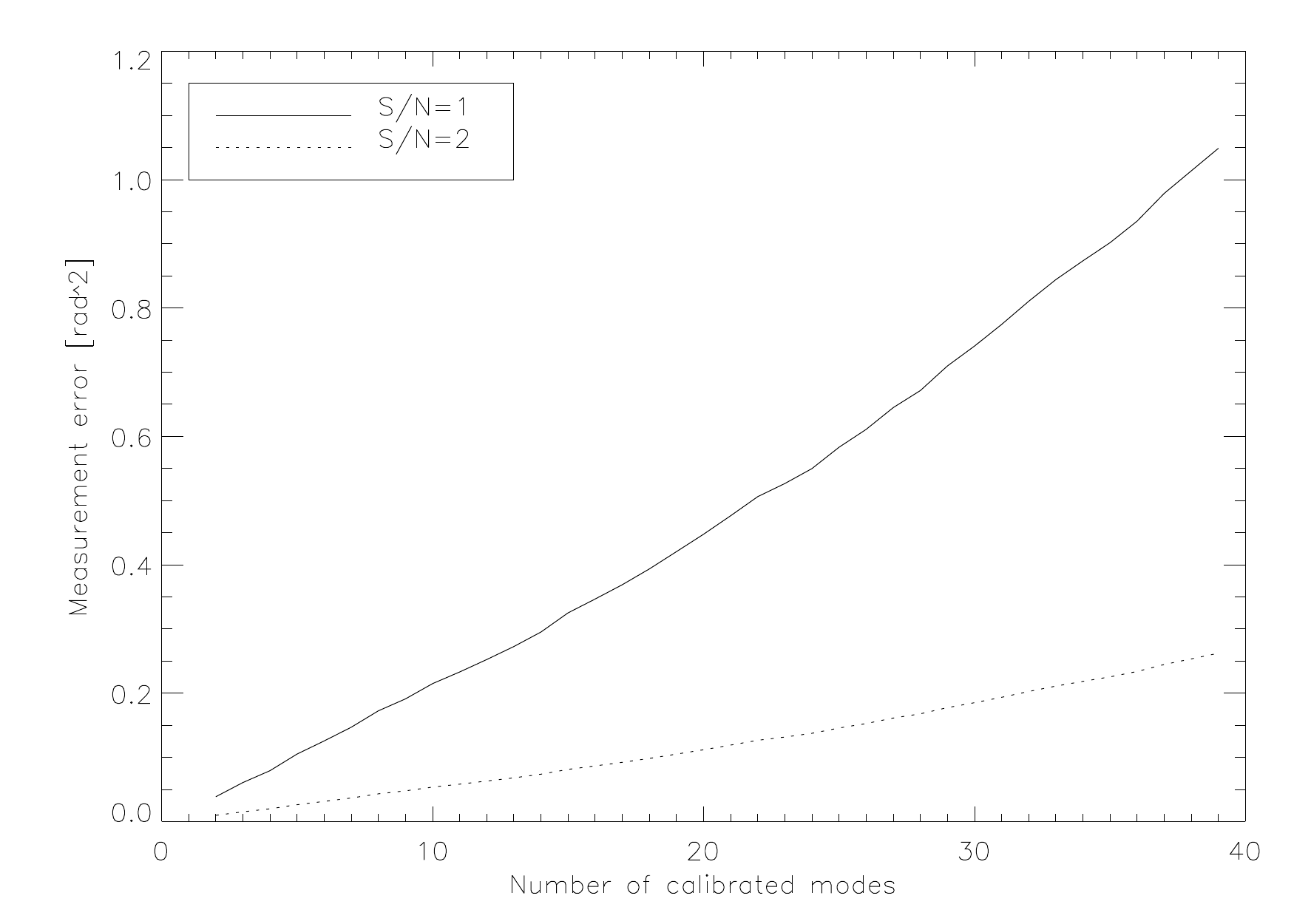}
   \caption{Left: dependence of the averaged modal cross talk coefficient on the
     number of modes ({\it solid line}). The dotted line is a $x^{-1}$ profile. Right: dependence of the measurement error on the number of calibrated
     modes for two values of S/N in the subapertures. The solid line denotes a
     S/N of 1 and the dotted line a S/N of 2.}
   \label{im:f24}
 \end{figure}
Fig.\ref{im:f24}, left panel, shows that the averaged modal cross talk coefficient falls off with the number of calibrated modes. The
best fit with a simple model yields a dependence for $N$ calibrated modes of
$N^{-1}$ as can be seen in Fig.\ref{im:f24}, right panel. Thus the total measured 
error due to modal cross talk in open loop can be derived to be,
\begin{equation}\sigma^2_{cr}=\sum_{i=1}^N<a_i^TR^TRa_i>=NE^{\prime
    2}\sum_{i=1}^N
  a_i^2\propto 0.3\left({D\over{r_0}}\right)^{5\over3}E_0^{\prime
    2}(1-N^{-{{\sqrt{3}}\over{2}}})\end{equation}
Here $E_0^2$ denotes the error for one calibrated mode only. Therefore the
modal cross talk will become independent of the number of modes for a higher number
of calibrated modes.
The benefit here is also that
the strength of the modal cross talk will be strongly reduced when we close the
loop because only the residuals of the corrected modes will give rise to
modal cross talk. \newline  
To find the best number of modes to calibrate, we have to include the
measurement error as well. This error depends on the S/N in the subapertures like
\begin{equation}\sigma_{meas}^2=tr[R^TR(S/N)^{-2}]. \end{equation}
The dependence of the trace of $R^TR$ on the number of calibrated modes is shown
in Fig.\ref{im:f24}, right panel. This trace is the sum over the squares of the entries of
the reconstruction matrix. Therefore a perfect linear dependence will arise
from a diagonal matrix $R^TR$. Modal cross talk is the reason for any
departure from this behavior. The departure is , however, very small. Thus we can
write the dependence as 
\begin{equation}\sigma_{meas}^2=tr[R^TR(S/N)^{-2}]\propto N\end{equation} 
 \begin{figure}[h!]
\centering
   \includegraphics[width=12cm]{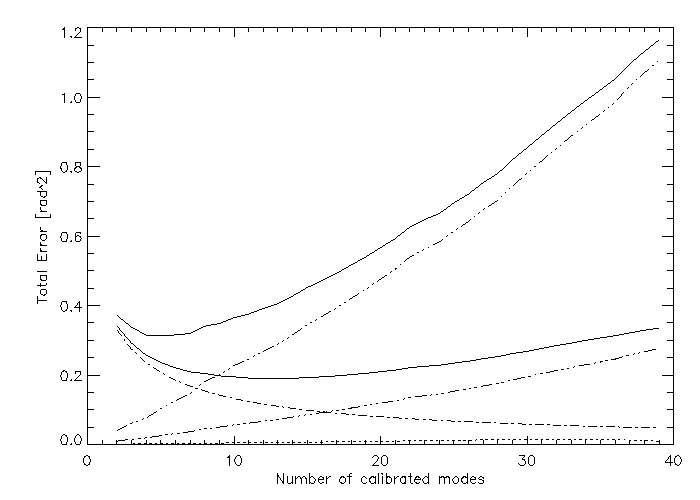}
   \caption{Different errors in dependence on the number of calibrated
     modes. The dotted line shows the aliasing error, the dashed line shows the error
     due to modal cross talk, the triple dot-dashed lines show the measurement
     error for S/N=1 and 2, the dash-dotted line shows the fitting error, and the solid
     lines show the total error for S/N of 1 and 2. }
   \label{im:f26}
 \end{figure}
 Fig.\ref{im:f26} compares these errors. The aliasing and modal
cross talk are of the same order and quite small. The measurement error is
shown for S/N=1 and 2. The errors due to modal cross talk and aliasing are small
compared to the others. The error , however, will decrease quadratically with
rising S/N. For S/N=1 the measurement error matches the fitting error at 8 calibrated
modes. The point of best correction (minimum of solid curve) is at 5 modes.
For a S/N of 2 the largest contribution arises from the fitting error for up to 16
modes. The best correction is given for 13 calibrated modes. These
curves are very helpful because we are able to change the number of modes we
use for reconstruction on the fly, adapting to the actual situation.
\subsection{Prediction of the On Sky Performance}
The errors presented in the last subsections were of static nature. To derive
the full error budged we need to include the temporal errors due to finite
loop bandwidth $ \sigma^2_{band} $ and time delay $ \sigma^2_{delay} $. These errors are given as \cite{Hardy1998}
\begin{eqnarray} 
\sigma^2_{band}&=&\left(f_G\over{f_S}\right)^{5\over{3}}\\
 \sigma^2_{delay}&=&28.4\left({f_{G}\tau_S}\right)^{5\over{3}}=28.4\left({{0.427v\tau_S}\over{r_0}}\right)^{5\over{3}}
\end{eqnarray}
with
\begin{eqnarray*}
  r_0&=&\,Coherence\,length\\
  f_{G}&=&\,Greenwood\,frequency\\ 
  \tau_S&=&\,Time\,delay\\
  v&=&\,Averaged\,wind\,speed\\
f_S&=&\,Loop\, bandwidth\,\approx 1/20\,loop\,speed.\end{eqnarray*} 
In the following the sum of the two temporal errors will be referred to as
$\sigma^2_{temp}$\newline
\begin{figure}[h!]
\centering
   \includegraphics[width=8cm]{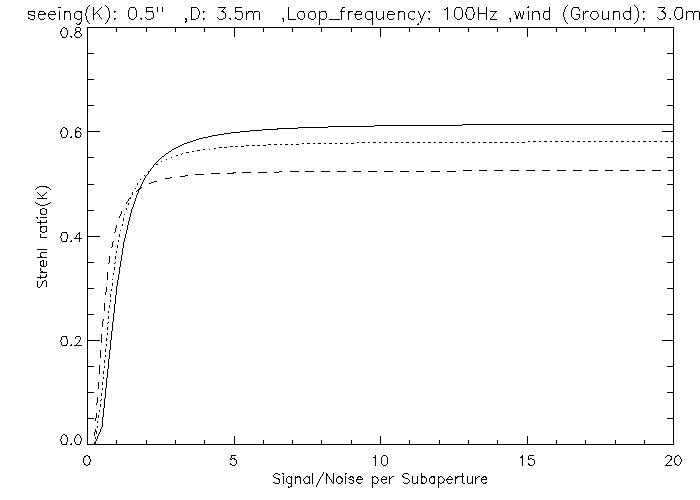}
\includegraphics[width=8cm]{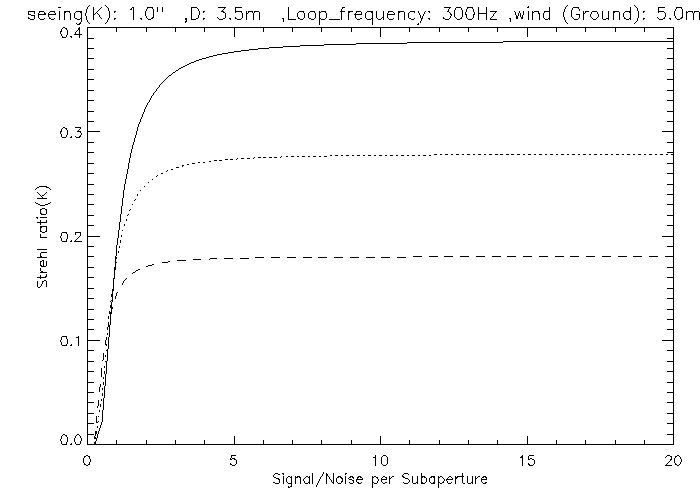}
   \caption{Left: predicted performance in K band for PYRAMIR on the sky for good
     seeing conditions. The three curves mark corrections with 12({\it dashed curve}),
     22({\it dotted curve}) and 32({\it solid curve}) modes.D denotes the
     telescope size, the wind speed is measured at ground level. Right: same as in the left panel, but for bad seeing conditions. Note
     that the bandwidth of the loop has been changed to 300 Hz. The curve
     styles are the same as in the left panel.}
   \label{im:f27}
 \end{figure}
To derive the
averaged wind speed from our measurements on ground level we used the 2/5 ratio between the
wind speed at a height of 200m and the wind speed on ground
level found on Paranal. This ratio numbers approximately 2. We adapted this factor into our
model. \newline
For the closed-loop application we also have to take into account that the
cross talk amplitude is reduced with respect to the open loop case due to the
correction by the system. Therefore the weighting of the cross talk is given
by the residuals from aliasing, temporal error and measurement error only,
\begin{equation}\sigma^2_{cr}\propto
  E_0^2(\sigma^2_{temp}+\sigma^2_{al}+\sigma^2_{meas})\end{equation}
The cross talk itself should occur on the right hand side also but has been
omitted because in CL it is much smaller than the other errors.

Now we can derive the expected Strehl values on sky as,
\begin{equation}SR=\exp[-(\sigma^2_{cross}+\sigma^2_{meas}+\sigma^2_{alias}+\sigma^2_{fit}+\sigma^2_{temp})]\end{equation}
 Fig.\ref{im:f27}, left panel, shows the performance for a seeing of 0.5'' in K-band and a wind speed
on ground level of
of 3 $ms^{-1}$ for a loop update rate of 100 Hz and 12, 22 and 32 modes corrected. A
Strehl ratio up to about 60\% is expected for bright stars. The limiting
S/N is about 0.25 per subaperture. This corresponds to a
stellar magnitude of 7.2 mag in K' with 20\% light on PYRAMIR.

Fig.\ref{im:f27}, right panel, shows the same for a seeing of 1'' in K-band a wind speed of 5 m/s and
a loop update rate of 300 Hz. Here we can still achieve about 39\% Strehl ratio. The
difference in performance between the number of modes in use is much more
significant. The limiting signal to noise ratio is about 0.25 per subaperture.
This corresponds to a magnitude of 6.0 mag in $K^\prime$ with 20\% light on PYRAMIR.
The reason the stars have to be so bright is the high sampling with 224
subapertures and the noise of 20$e^{-}$ rms per pixel.
\section{Conclusions}\label{sec:conclusion}
In this article we presented laboratory measurements performed with the pyramid
wavefront sensor PYRAMIR.  After a short introduction into the history of PWFS
and their working principles, we presented the PYRAMIR system as infrared
PWFS and explained the details of the system especially the possible read out
modes of the detector. This can reach a speed of about 300 Hz with a RON of
20 $e^-$ rms.  The calibration procedure of the system was described in
detail. The pitfalls of TT-calibration were discussed in detail leading to the
conclusion that a static part in TT will reduce the limiting magnitude and
enhance TT-jitter and cannot be tolerated. Therefore we included a static
TT-part in the bias pattern of the deformable mirror (dmBias) to perfectly
center the beam. Also it has to be guaranteed that the amplitudes of
calibration are the same in both axis in order to gain similar performance in
both directions.\newline Some of the fundamental sources of reduced
performance were examined. We found that the amount of light diffracted out of
the pupil images is 50\% for a flat wavefront decreasing nonlinearly up to 2
$rad$ wavefront error where it becomes almost stable at 20\%. A comparison
with simulations shows that in our case this loss results from diffraction at
the pyramid edges and imperfections in the optics. The latter outweighing for
aberrations stronger than about 1.5 $rad$.\newline The effects of a Gaussian
illumination of the pupils, as well as the effect of an extended calibration
light source, were investigated.  Both have strong influence on the sensitivity
of the system and should be avoided.  On the other hand an extended target
during the measurement does only slightly effect the performance. Thus the
calibration light source should be as point like as possible for all
applications.\newline The effect of RON on the limiting magnitude of the guide
star has been widely discussed. In the case of infrared detectors this noise
is quite high. In our case 20 $e^-$ rms per pixel. This reduces the limiting magnitude
by 5 mag.  We investigated into the best possible mode set between the
eigen modes of PYRAMIR, of the DM and the KL modes. The last two mode sets
seem promising for the use on-sky. The latter has a larger linear regime
than the eigen modes of the DM. The difference is
small and, therefore, the correction error will not vary by much. Still the larger
linear regime might help to close the loop under bad seeing conditions. \newline We
tested the best treatment of noncommon-path aberrations by applying
artificial modes to the DM. The best treatment turned out to direct these
aberrations completely into the path of the sensor. A possible better way
using two calibrations, one for modes with positive amplitude one for those
with negative amplitude was proposed to reduce the error of the mode with
static aberrations, but put aside due to the fact that it was not running
stably during the testing on-sky.\newline The importance of a small
calibration amplitude to minimize the resulting reconstruction error was
shown. \newline After the measurement of these fundamental properties, the
subject of the best number of modes to calibrate was addressed. To solve this
we investigated into the behavior of modal cross talk, aliasing, and
measurement error dependence on the number of modes calibrated. We found
that the averaged aliasing coefficient varies only slightly with the number of modes
whereas the average modal cross talk coefficient decreases inversely linear with this
number and the measurement error rises linearly with this number. Including
the (theoretical) contribution of the KL modes in the wavefront error on sky
we could show that the contribution of aliasing rises like $N^{(2-\sqrt{3})\over 2}$; the
contribution of cross talk becomes constant for larger $N$ and the measurement
error rises linearly with $N$. In CL only the error due to modal cross talk
changes with respect to the open loop measurement. It will decrease because
the modes that are corrected will contribute less to the modal cross talk
than in open loop. The error of the residual wavefront decreases like
$N^{-{\sqrt{3}\over 2}}$. Altogether we could show that at the border of the limiting
magnitude the fitting error surpasses all other errors for a low number of
corrected modes but will be overpowered by the measurement error at about the
place of the optimum number of modes to be corrected. For the PYRAMIR system
this is 5 modes.  The other errors will become important for even slightly
brighter stars. Here cross talk and aliasing error are almost identical in
strength.  In the case of noncommon-path aberrations in the system the modal
cross talk and aliasing errors will rise. Modal cross talk increases linearly, aliasing
error stays constant until the static aberration reaches the border of the
linear regime of the sensor, then it increases nonlinearly. \newline From the
entire error-budged we could predict the performance on-sky for various seeing
conditions. For a seeing of 1'' and a wind speed on ground level of 5$ms^{-1}$ we can achieve a 39\%
Strehl ratio in $K^\prime$-band, but we have to run at about maximum frame rate (300
Hz). The limiting S/N per subaperture on the detector is
about 0.25 or 6.0 mag in $K^\prime$.  For good seeing conditions and a moderate loop
band width, the predicted Strehl ratio will be about 60\% for bright stars.
Again the limiting S/No per subaperture on the detector is
about 0.25 or 7.2 mag in $K^\prime$.

\end{document}